\documentclass[
reprint,
 showpacs,
 preprintnumbers,
 amsmath,
 amssymb,
 pra,
]{revtex4-1}

\usepackage{graphicx}
\usepackage{dcolumn}
\usepackage{bm}

\begin{document}


\title{Classical sources of non-classical physics:\\the case of linear superposition}

\author{Ghenadie N. Mardari}
 \email{gmardari@gmail.com}

\author{James A. Greenwood}
\affiliation{
 Open Worlds Research\\
 Baltimore, Maryland, USA
}

\date{\today}

\begin{abstract}
Classical linear wave superposition produces the appearance of
interference. This observation can be interpreted in two equivalent
ways: one can assume that interference is an illusion because input
components remain unperturbed, or that interference is real and
input components undergo energy redistribution. Both interpretations
entail the same observable consequences at the macroscopic level,
but the first approach is considerably more popular. This preference
was established before the emergence of quantum mechanics.
Unfortunately, it requires a non-classical underlying mechanism and
fails to explain well-known microscopic observations. Classical
physics appears to collapse at the quantum level. On the other hand,
quantum superposition can be described as a classical process if the
second alternative is adopted. The gap between classical mechanics
and quantum mechanics is an interpretive problem.
\end{abstract}


\maketitle


\section{Introduction}

When two waves overlap, they merge into a single formation and the
principle of superposition applies: at every point, the net
amplitude is equal to the vector \emph{sum} of the input components.
In contrast, the energy of a wave is proportional to the
\emph{square} of its net amplitude. This means that the net state
acquires a surplus of energy at every point of constructive
interference. It also loses a complementary amount at every point of
destructive interference. The total amount of energy remains
constant, but the problem is to explain the underlying physical
mechanism. Does energy flow locally from the areas with destructive
interference to the areas with constructive interference, or does it
simply vanish and appear independently at each point? The
observation of redistribution cannot be dismissed as an illusion,
because useful work can be extracted from net states, according to
their content. In particular, it is impossible to extract energy
from a null region, but the extra energy is readily available in the
areas with constructive interference.

\begin{figure}
\includegraphics{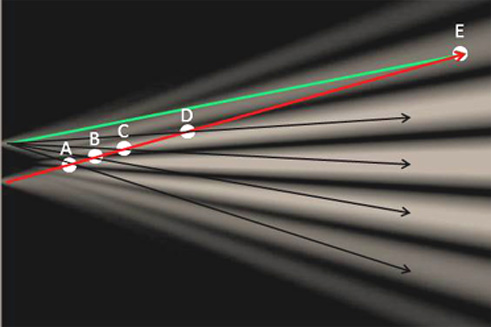}
\caption{\label{fig:rays} (Color online) \textbf{Interference without
consequence.} The net state of any point in the interference volume
(\emph{e.g.}, E) is determined by the sum of component amplitude
vectors. The crucial parameter is phase difference. In the case of
coherent beams, it is calculated by comparing the shortest path
lengths from each source to the target. It is irrelevant how many
rays from the first source intersect with a ray from the second
source (points A -- D) before it reaches the target. For this
reason, it is tempting to assume that waves propagate through each
other unperturbed. Though, as shown below, this assumption is
unnecessary and entails a violation of the principles of classical
physics.}
\end{figure}

This problem is aptly captured by the question: do waves go through
each other unperturbed? The typical response is to adopt a positive
answer. If one takes a moment to observe the propagation of water
waves in a ripple tank, it is very hard to deny the appearance that
waves are transparent to each other. Similarly, human ears have no
trouble distinguishing simultaneous sounds from different sources,
just like the multi-media devices that isolate electromagnetic
signals with high fidelity from complex input mixtures. If that is not enough,
rational arguments can be used to support this position as well. For
example, the principle of superposition applies to any point in the
interference volume of two light beams. Yet, a ray from the first
source must collide with billions of rays from the second source
before it reaches a region of interest (Fig.~\ref{fig:rays}). As it is
known, the net state at any point is calculated with utmost accuracy
just by taking into account the distances between the target and
each source. The slightest perturbation would have been amplified at
the numerous intermediate points, falsifying the underlying principle.
In short, it seems necessary to assume that waves go
through each other unperturbed, without really interacting with each
other.

The principle of non-interaction of waves is widely taken for
granted as the correct interpretation of linear classical
superposition. It is omnipresent in the literature at any level,
from introductory textbooks to high profile research reports.
Unfortunately, it has a major interpretive shortcoming. If
interference is assumed to be an illusion, then energy
redistribution (which is not an illusion) has to be described as a
non-local process. According to this approach, it seems unavoidable
to conclude that classical interference is fundamentally
non-classical. Yet, this is not the only valid interpretation at our
disposal. The same predictions would be obtained if we assumed that
energy redistribution was a local process. For example, Richard
Feynman dedicated a whole chapter to this problem in his
\emph{Lectures on Physics} \cite{feyn64}. He admitted that local
energy conservation sounded strange in some contexts, but he also
emphasized its necessity. The theory of relativity and the principle
of momentum conservation would be in trouble as ontological models,
if energy redistribution was non-local in the physical world.

To some readers, Feynman's conclusions might sound
counter-intuitive. Perhaps, when two identical waves overlap, it
might be hard to tell if they go through each other or bounce back.
We get identical outputs in either scenario. Yet, more often than
not, overlapping waves are not perfectly identical. When a tall wave
intersects with a short wave, the two of them keep on going in the
original directions, as if they never met each other. Thus, it seems
inappropriate to suggest that the waves \emph{do not go} through
each other unperturbed. Nevertheless, rigorous quantitative analysis
suggests that the two models are indeed equivalent. For example,
Dowling and Gea-Banacloche analyzed the behavior of intersecting
light beams under the assumption that they bounce off, instead of
going through each other at the microscopic level \cite{dow92}. They
found that the macroscopic predictions are identical in both cases,
even if the input beams are distinguishable by amplitude, frequency,
or polarization. Output beams are quantitatively similar to the
input components, but cannot be assumed to be qualitatively
identical because of the possibility of symmetric energy exchanges
during interference.

As it turns out, there are three major interpretations of wave
interference that are experimentally indistinguishable. One can
assume that waves go through each other unperturbed, or that they
undergo specular reflection, or finally that input waves merge into
a single net state that mimics the appearance of transparent passage
in its evolution. The latter seems particularly vulnerable to
objections. If two unequal waves are truly able to merge into a
single net state, then why don't we get two outputs with equal
amplitudes? Why does it seem that the taller wave maintains its
original shape and direction after overlap? The answer is that this
objection and the ones invoked earlier in this text are based on
superficial observations. Wave behavior is not predicted by
macroscopic qualities, such as ``shape'', but rather by the analysis
of microscopic processes in accordance with Huygens' Principle. The
net state at every new frontline is calculated by adding up the
effect of wavelets from every point of a previous frontline. Whether
we apply this method at the level of net states or independent
components, the final predictions are exactly the same in all three
models. Indeed, we would have to deal with a mathematical
contradiction if this was not the case. The relevant distinction
between the three scenarios is only found in the domain of
qualitative considerations. The ideas that waves can ``go through''
or ``bounce off'' each other are based on particle models of
propagation. Yet, mechanical waves do not transport particulate
matter. They only carry momentum from one region to the next in
elastic media. Thus, only the third approach is compatible with the
classical notion of a mechanical wave, and it is also the one that
does not entail any complication with regard to energy
redistribution.

Modern physics has two incompatible branches: classical mechanics
and quantum mechanics. The usual assumption is that classical
mechanics is easy to interpret, while quantum mechanics is not. We
wish to suggest that the interpretive problems of classical
mechanics have been underestimated, and that wave interference in
particular has been predominantly tackled with a model that is
qualitatively non-classical. This is especially relevant for the
debate about the boundary between classical phenomena and quantum
phenomena. A growing number of scientific conferences have been
recently devoted to this issue. After attending some of them
\cite{qtrf3, feynfest, fpp4, wap2, qtrf4, wap3, qtrf5, wap4}, we
were surprised by the sheer number of arguments in favor of the
formal compatibility between the two branches. This inspired us to
inquire if the gap between classical mechanics and quantum mechanics
is truly ontological. Would we get identical conclusions, if
classical phenomena were subjected to the same level of scrutiny as
quantum phenomena with regard to their qualitative implications?
This essays has three goals: 1) to explain the non-classical essence
of current approaches to wave interference; 2) to share our findings
regarding the quantitative and the experimental equivalence of
alternative approaches to this phenomenon; and 3) to outline the
relevance of these conclusions for the understanding of quantum
phenomena.

\section{The hidden dimensions of linear superposition}

\begin{figure*}
\includegraphics{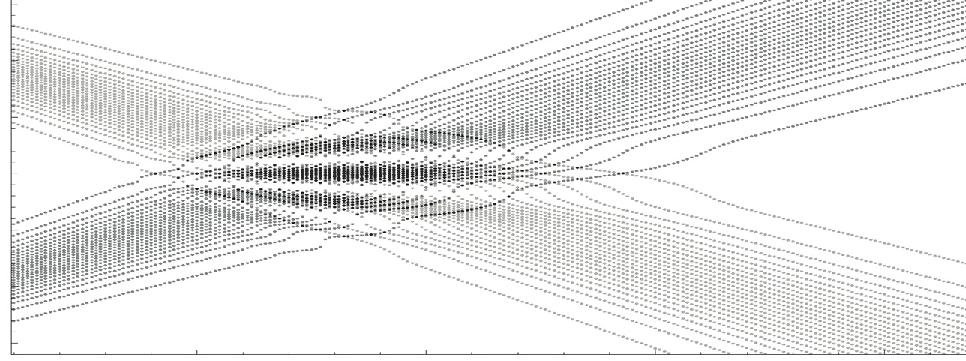}
\caption{\label{fig:beams} \textbf{Visual aid for the analysis of beam
superposition.} The distribution of energy in the cross-section of a
beam is represented by 25 dots, where each dot marks the center of a
volume containing 4\% of the beam's total energy. As a result, the
concentration of dots in each vertical column serves as a visual
indicator of energy density. Given the presence of two beams, there
are 50 dots in all for each z coordinate. The evolution of the
observable states of energy is calculated as a function of distance
from the source. This model is not informed by any ontological
assumption. It simply creates a panoramic view of all the stages of
superposition between the beams, given their calculated amplitudes
and phase difference at various points. Hence, when we are trying to
decide if the beams go through each other unperturbed, or if they
actually redistribute their energy during interference, this is the
kind of pattern that we are trying to interpret.}
\end{figure*}

Non-linear wave superposition is a process with clear physical
consequences: input components are transformed and become
unrecognizable at the output. In contrast, linear wave superposition
is very ambiguous. On the one hand, output states are similar to
input states, as if no interaction ever takes place. On the other
hand, interference is observable in the area of overlap
(Fig.~\ref{fig:beams}).
Is the similarity between inputs and outputs illusory, or is
interference just a misleading appearance? The main challenge is to
explain what happens in the coincidence volume . Why do we see
energy redistribution? Is it because detectors sum up the effect of
independent components, as they pass through each other unperturbed,
or is it because the two components really interact and change their
physical properties at the microscopic level? From a quantitative
point of view, it is far more convenient to assume that waves do not
really interfere. The outcome is the same in both scenarios, but the
calculations are much simpler in this case. Though, Brownian motion
can also be ignored at some levels of analysis where it is
inconsequential. That does not change anything about its status as a
real phenomenon. Why should interference be different? A possible
answer is that Brownian motion has independent effects that cannot
be explained otherwise. The same cannot be said about interference:
every known effect of this phenomenon can be reproduced by assuming
that superposed components never interact. What is there to be lost,
if the reality of interference is denied? Our reply, as explained
below, can be summarized in two words: \emph{classical mechanics}.

\subsection{Classical waves cannot propagate \\ through each other}

\begin{figure*}
\includegraphics{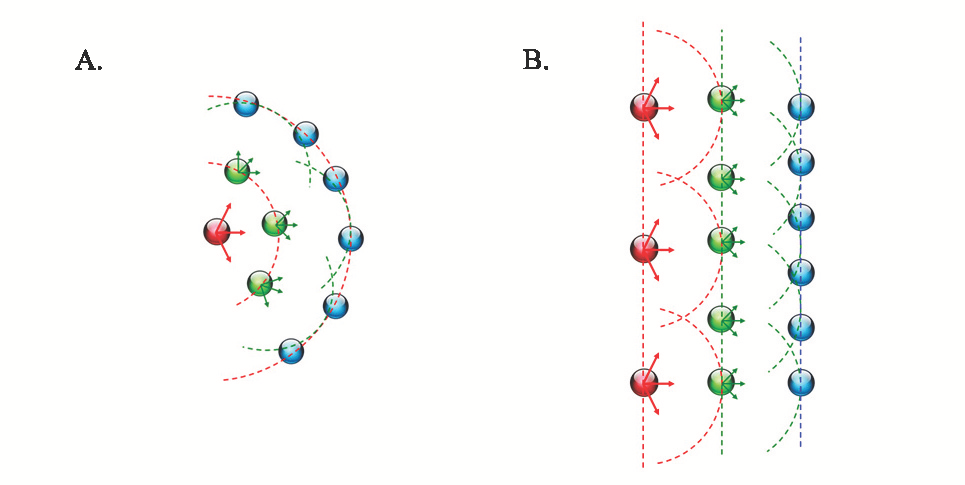}
\caption{\label{fig:wavelet}(Color online) \textbf{A toy model for
Huygens' Principle.} Classical molecular media can be mimicked with
a macroscopic construct, in which identical solid balls are
interconnected in a symmetric mesh with elastic springs. The whole
structure is under tension, due to the compressive effect of
gravity. At rest, there is a state of equilibrium in which the balls
are assumed to be equidistant from each other. (A) If a single ball
is suddenly displaced from its point of equilibrium, it will intrude
on the adjacent space in the direction of action. All of the
immediate neighbors from the point of maximal intrusion will
experience a net radial outward force. As they move away from this
source of compression, they must produce a pattern of motion that
can be described as a spherical wavelet (at least at the early stage
of the process). These displaced balls are going to have a similar
effect on their own neighbors. Yet, given that a larger number of
balls act at the same time, their impact will add up constructively
or destructively at different points. The net effect will be a
macroscopic spherical front of constructive interference. (B) If a
large number of balls from the same plane are displaced
simultaneously in the same direction, they will produce wavelets
like in the previous example. Yet, this time their effect will add
up to a front of constructive interference that resembles a plane
wave, rather than a spherical wave. Thus, Huygens' Principle
captures very closely the dynamics of perturbation of molecular
media. At the macroscopic level, the distance between any two
molecules is negligible. Accordingly, every point on a wave-front
can be treated as a source of secondary wavelets. Note that only the
points on the front of constructive interference are relevant. The
regions with destructive interference contain entities with zero net
displacement. If the latter do not move, then they cannot produce
compression. Therefore, they cannot act as sources of wavelets.}
\end{figure*}

A classical mechanical wave is a pattern of oscillation that is
produced by the relative displacement of small regions in an elastic
medium. The cause of this phenomenon is the tendency of excited
volumes to return to their state of equilibrium, rather than to run
away. In other words, mechanical waves do not transport matter. When
they run, it is because the state of motion of one region is
transferred to the next. For example, consider the effect of a
membrane on a gaseous medium (\emph{e.g.}, air). In a toy model for
this process (Fig.~\ref{fig:wavelet}), atmospheric particles can be
replaced by solid balls, interconnected by elastic springs that can
transfer momentum from one region to the next without losses.
Whenever one molecule is pushed by the action of the membrane, it
gets closer than normal to its adjacent particles. As a result, the
balance between the forces of these entities is disrupted. The
action of the displaced molecule on its neighbors is temporarily
stronger than the action of the rest of the medium. These adjacent
particles end up being displaced as well. The outcome is a close
analogue to a half-spherical wavelet, as captured by the Huygens
principle. Applying this description to each molecule that is
displaced by the first one, we can follow the same principle in
order to arrive at the description of a macroscopic half-spherical
wave-front. If the active size of the membrane is increased, many
molecules are displaced simultaneously in the same plane. The
macroscopic effect of their wavelets is to produce an interference
front that is closer to the plane wave approximation. As explained
by Huygens \cite{huyg12}, the final shape and direction of a
macroscopic front is determined by the amplitudes and the relative
phases of microscopic wavelets. The latter may overlap in many
directions, but the wave-front develops in the direction where they
add up constructively. A particle will only produce wavelets if it
moves, and it can only move if it receives a net momentum in a given
direction. Hence, a ``wave front'' is really a ``constructive
interference front''.

\begin{figure*}
\includegraphics{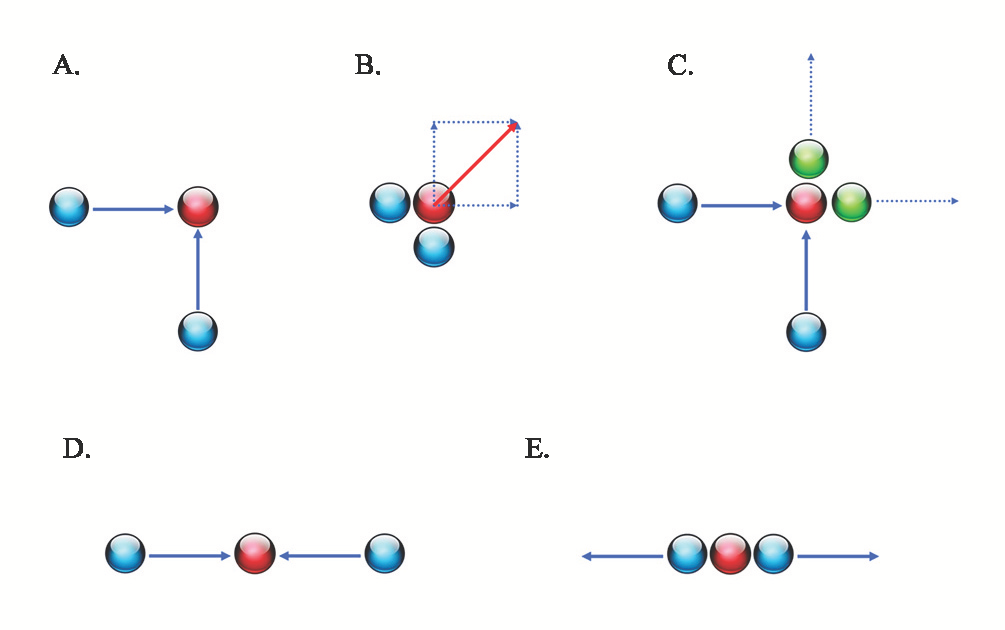}
\caption{\label{fig:collide} (Color online) \textbf{Momentum transfer
during linear wave superposition.} Macroscopic waves are reducible
to microscopic instances of momentum transfer in classical media. In
the same vein, superposition can be analyzed in terms of
simultaneous elastic collisions. (A) Two blue balls exercise
pressure on the same red ball from orthogonal directions. If the
upward ball acted alone, the red ball would be displaced vertically.
Similarly, the red ball would move to the right, if the other blue
ball acted alone. Yet, the red ball cannot move in two directions at
the same time. (B) The only possible way for it to move is in the
direction of the net force vector. Moreover, the input components
cannot be physically preserved in the net state of motion. The ball
is simply moving. The same motion could be produced by two balls
acting from other angles, or by any other number of balls. (C) Two
green balls appear to carry forward the exact momentum of
corresponding blue balls. Their motion is quantitatively identical
whether blue balls act one by one or at the same time. Is it also
qualitatively identical? The red ball can only move diagonally in
the case of simultaneous action. Ergo, the direction of the green
balls is determined by their position relative to the moving red
ball. If the starting arrangement was not symmetric, the green balls
would carry the same total momentum in different directions. (D-E)
This is a different context, in which two blue balls act
simultaneously from opposite sides. The net force acting on the red
ball is equal to zero, so that it does not move. Instead, it acts as
a rigid wave-breaker. The blue balls appear to transfer their
momentum to each other, but this is physically impossible if the red
ball cannot move. It is at least equally plausible that they bounce
back with their own momentum, or that the two inputs merge into a
single state and become indistinguishable. Ergo, it is no longer
possible to determine unique origins for the output momentum states.
These two examples suggest that input components must be reflected
from the areas with destructive interference and become
indistinguishable in the areas with constructive interference, as
these types of interactions are repeated numerous times during
linear wave superposition. In order to avoid this conclusion, one
might have to assume that real classical entities move in several
directions at the same time, and also transfer real momentum without
moving.}
\end{figure*}

When two different wavelets overlap on a single point, it means that
one particle is the recipient of action from different directions.
For simplicity, this process can be illustrated in terms of elastic
collisions between macroscopic balls (Fig.~\ref{fig:collide}A-B). If a
ball is hit horizontally from the left, it must be displaced to the
right. If it is hit from below, it should end up moving upwards. The
only way to claim that input momentum can ``go through'' is to show
that the receiving particle is able to transfer it \emph{by moving}
accordingly. However, classical mechanical entities cannot move in
two directions at the same time! When several sources exercise their
action simultaneously, the target can only move in the net
direction, in this case diagonally. (Hence, the corresponding
particle in the medium will only be able to initiate a wavelet in
the net direction). If two identical balls are placed in the
appropriate configuration after the impact, they may end up moving
in the original direction of the inputs (Fig.~\ref{fig:collide}C). Yet,
the role of the intervening carrier, which merges all the input
units of action into a single physical motion, is to replace the
input states, effectively destroying them. The latter can be
recreated afterwards, but they cannot be preserved. Similarly, when
two balls with equal momentum arrive simultaneously from opposite
directions (Fig.~\ref{fig:collide}D-E), the central ball cannot be
displaced at all.
If it does not move, then it cannot carry momentum of any sort in
any direction. Its physical function is to work as a ``wave
breaker'', like a rigid wall. From a quantitative point of view, it
makes no difference whether we assume that two identical balls
recoil from each other or pass through - the outcome is the same.
From a qualitative point of view, things are very different. As long
as the principles of classical mechanics are assumed to be at work,
only the first alternative is plausible. For this reason, the action
of air molecules cannot be assumed to propagate ``through'' the
areas of destructive interference. Their momentum has to be
redirected into the areas with constructive interference. As a
result, the molecules from the regions with constructive
interference must oscillate with higher amplitudes. They carry the
summed momentum of particles that happen to push directly through,
and of the particles that recoiled from adjacent regions. In
contrast, the molecules from the areas with destructive interference
cannot move at all and their amplitude of motion is null. This is an
intuitive explanation of the known properties of macroscopic
wavefronts, as produced by microscopic wavelets.

Huygens' model implies that a wave is nothing but a state of
constructive interference on a medium. To describe the energy of a
macroscopic wave is to describe the area of constructive addition of
microscopic wavelets. For example, the diffraction angle of a wave
can be changed by inducing phase delays between wavelets
\cite{bak87}. When a wave is focused, all the billions of wavelets
can be assumed to have real amplitudes in the original (unfocused)
direction. It is only the front of constructive interference that
converges on a point. Nevertheless, the common sense description is
that \emph{the wave} converges and that the energy in the focal
point is really all the energy available. The wave happens where the
medium oscillates. To speak of energy outside the detectable wave is
both counterintuitive and impractical. If the wave is not assumed to
correspond to the observable oscillations, energy must be described
as if it was spread over the entire medium and then it becomes
impossible to describe what propagates, where, and how. Indeed,
there are no debates in the scientific community (to the best of our
knowledge) about this aspect of the nature of single waves. The
problems only emerge when several coherent waves overlap. In this
case, typical patterns emerge in the form of interference fringes,
but the waves appear to return to their original shapes after
overlap, at least in some cases. For reasons that do not concern us
here, interference is suddenly interpreted as a process with
unperturbed input energy, even in the destructive null zones where
no oscillations are detectable. This is a major inconsistency. A
single wave is a process of superposition between numerous
microscopic wavelets. Two-wave superposition is the same process,
multiplied by two. Yet, single-wave energy is presumed to be
localized exclusively in the observable oscillations, while
double-wave energy is not. Moreover, non-interference implies that
elementary particles in the medium can move in several directions at
the same time, and even transfer momentum without moving at all!
That is a clear departure from the principles that define Newtonian
physics.

The motion of any physical entity may be represented by a vector
during formal analysis. Any ``real'' vector, in turn, is equivalent
to the sum of two or more ``virtual'' vectors that add up to the
same net state. Sometimes, these components are physically
significant, but their virtual nature is self-evident. For example,
if two football players kick the same ball at the same time, the
action of each of them can be represented by a dedicated vector, but
the ball can only move in the net direction. Its displacement is
captured by a ``real'' vector, while its hypothetical components
(had it been struck by either player alone) must be represented by
``virtual'' vectors. The same relationships are found in the
behavior of a medium when two waves become superposed. The net state
is detectable, but the components are not. Yet, here we get the
unique standard operating procedure, passed down from one generation
of scientists to the next, to assume that unobservable input
components are ``real'', while the observable net states are
``virtual''. Regardless of the practical advantages of this
preference, it comes with the heavy interpretive toll that was
described in the previous paragraph. The ontology of this approach
is non-classical and the analysis of wave behavior is conceptually
inconsistent.

A possible objection to this argument concerns the relevance of real
elastic media for a general approach to classical waves, given that
light waves appear to propagate without a medium. Indeed, some
textbooks maintain that electromagnetic waves are not compatible --
in any physical sense -- with the Huygens postulate, even though it
works with unrivalled accuracy \footnote{For example, M. Schwartz is
very explicit on this point, in his \emph{Principle of
Electrodynamics} (Dover, 1983): ``We are all qualitatively familiar
with the fact that a plane wave of light passing through a small
hole in a wall exhibits a remarkable interference pattern on the far
side. This pattern is generally explained in terms of the so-called
Huygen's principle, which tells us to consider each point on a
wavefront as a new source of radiation and add the `radiation' from
all of the new `sources' together. Physically, this makes \emph{no}
sense at all. Light does not emit light; only accelerating charges
emit light. Thus we will begin by throwing out Huygen's principle
completely; later we will see that it actually does give the right
answer for the wrong reasons.'' (p. 293).}. In particular, they
question the idea that light can be treated as a source of light at
every point of its wave-front \cite{schw83}. The relevant concern
here is that electromagnetic radiation can only be produced by
accelerated charges. If light is made of photons without charge that
move at a constant speed, it should be impossible to generate new
forms of radiation. On closer inspection, this problem appears to be
based on a misunderstanding. Huygens' postulate does not entail that
waves produce wavelets. Instead, it redefines the waves, by
describing them as the net state of a totality of wavelets. A wave
can be described \emph{either} as a running perturbation with a
specific shape, \emph{or} as a process of sequential generation of
secondary wavelets that add up to the same shape. Hence, only
wavelets produce wavelets and the starting conditions (be they
mechanical oscillators or accelerated charges) only serve to explain
the origins of this process. According to the most common
interpretation, electromagnetic waves propagate by generating
component fields. Changing electric fields induce changing magnetic
fields that induce changing electric fields and so on. At least
formally, these fields can operate like the wavelets of other types
of classical waves \cite{hecht}, because they are assumed to be
constantly created locally at every new wave-front. More
importantly, the same considerations about the net state versus
component states apply to this case. For example, when two electric
fields act on the same point, the resulting magnetic field will have
detectable effects only in the net state. This aspect will be
described in greater detail in the following sections. In short,
light waves are both quantitatively and qualitatively compatible
with the properties of other types of classical waves.

\subsection{Superposition entails energy redistribution}

Let us now consider the quantitative aspects of linear
superposition. When two laser projections overlap, the net
distribution of energy in the cross-section may be determined by one
of two rules. If the beams are mutually coherent, irradiance is
proportional to the square of the vector sum of their electric field
amplitudes at any point:
\begin{equation}
I=k(\bm{A}_{1}+\bm{A}_{2})^{2} \label{eq1}
\end{equation}
where $k=\epsilon_{0}c$. If the beams are incoherent, it is the sum
of the squared amplitudes that gives correct predictions:
\begin{equation}
I=k(A_{1}^{2}+A_{2}^{2})~. \label{eq2}
\end{equation}
Given that
\begin{equation}
(\bm{A}_{1}+\bm{A}_{2})^{2}=A_{1}^{2}+A_{2}^{2}+2A_{1}A_{2}\cos\theta
\label{eq3}
\end{equation} for any two vectors, the difference
between equation (1) and equation (2) reduces to
\begin{equation}
2kA_{1}A_{2}\cos\theta~, \label{eq4}
\end{equation}
where $\theta$, in this case, corresponds to the phase delay between
the two coherent wave patterns. Expression (\ref{eq4}) is generally
known as the ``interference term''. In practice, the rules
(\ref{eq1}) and (\ref{eq2}) account for the observed presence or
absence of interference fringes when multiple beams overlap, and
their application is straightforward. The difficulty is to explain
the meaning of the difference between them. Is there something
physical behind the interference term (\ref{eq4}), or is it just a
product of mathematical manipulation?

According to the left side of the equation (\ref{eq3}), we must
assume that the local state of two component fields is sufficient to
explain the observed amount of irradiance. The two fields do not
interact and their individual amplitudes remain constant over time.
Yet, their joint action can generate the appearances of fringes,
because of the varying phase angle between their amplitude vectors
at different points in the cross-plane. When the amplitudes point in
the same direction, we could assume that they add up. When they
point in opposite directions, they should similarly cancel each
other out. This relationship can only be stable over time for
mutually coherent beams. Hence, there is a plausible explanation for
the difference between the cases that require equation (\ref{eq1})
and the ones that require equation (\ref{eq2}). In contrast, the
right side of equation (\ref{eq3}) suggests a radically different
scenario. Instead of vector addition, we have the sum of two scalar
values. These individual components add up to a common state at
every stage of the process, just like in the case of incoherent
superposition. For this reason, the local properties of undisturbed
beams cannot explain the emergence of fringes. The latter must be
described by assuming the existence of a process of energy exchange
between adjacent regions, as captured by the interference term
(\ref{eq4}). As a corollary, we have two incompatible stories that
happen to be mathematically equivalent.

The standard interpretation of interference does not involve
redistribution. When crests overlap with troughs, they are presumed
to cancel out each other's effect. When crests overlap with crests,
the amplitudes are expected to resonate. This effect is somewhat
similar to that of two horses pulling a cart. If they pull in
opposite directions, there is no net displacement. If they pull in
the same direction, their force adds up. Unfortunately, this analogy
does not work if energy is taken into account. When horses pull in
opposite directions, a lot of energy is dissipated in the form of
heat. When waves are out of phase, no heat is released. Instead, the
waves are supposed to keep on propagating through each other.
Furthermore, when two horses pull in the same direction, it is
impossible to get more than two horse power. Yet, energy is supposed
to multiply when two wave crests overlap, because the linear
summation rule applies to amplitudes. For example, constructive
interference between two beams with equal amplitudes
$(\bm{A}_{1}=\bm{A}_{2})$ produces an irradiance that is:
\begin{equation}
I=k(2\bm{A}_{1})^{2}=4A_{1}^{2}k~. \label{eq5}
\end{equation}
This is twice as much as the irradiance of two incoherent beams in
superposition:
\begin{equation}
I=k(A_{1}^{2}+A_{1}^{2})=2A_{1}^{2}k~. \label{eq6}
\end{equation}
It makes sense to assume that incoherent beams cannot resonate.
Perhaps, their joint energy is lower because their oscillations
``average out'' somehow? The answer has to be negative because the
irradiance of incoherent beams in superposition is equal to the sum
of their individual irradiance, measured separately. If the
interaction of incoherent beams were to ``average'' something out,
the net state should be less than the sum of input components. Since
this is not the case, there is no local ``reserve'' for the extra
energy in a bright fringe, and an external source must be
identified. Amplitude vectors may cancel out and add up in various
ways, but energy is supposed to be a conserved scalar quantity.
Either it becomes observable in one spot when it is unobservable in
another, due to some sort of non-local transaction, or it has to be
physically redistributed by local means. Yet, the whole point of
amplitude summation was to avoid the conclusion of energy
redistribution. Therefore, energy modulation has to be treated as an
illusory but natural consequence of amplitude summation, according
to this approach, such that the appearance of overall energy
conservation during interference is merely a coincidence.

\begin{figure*}
\includegraphics{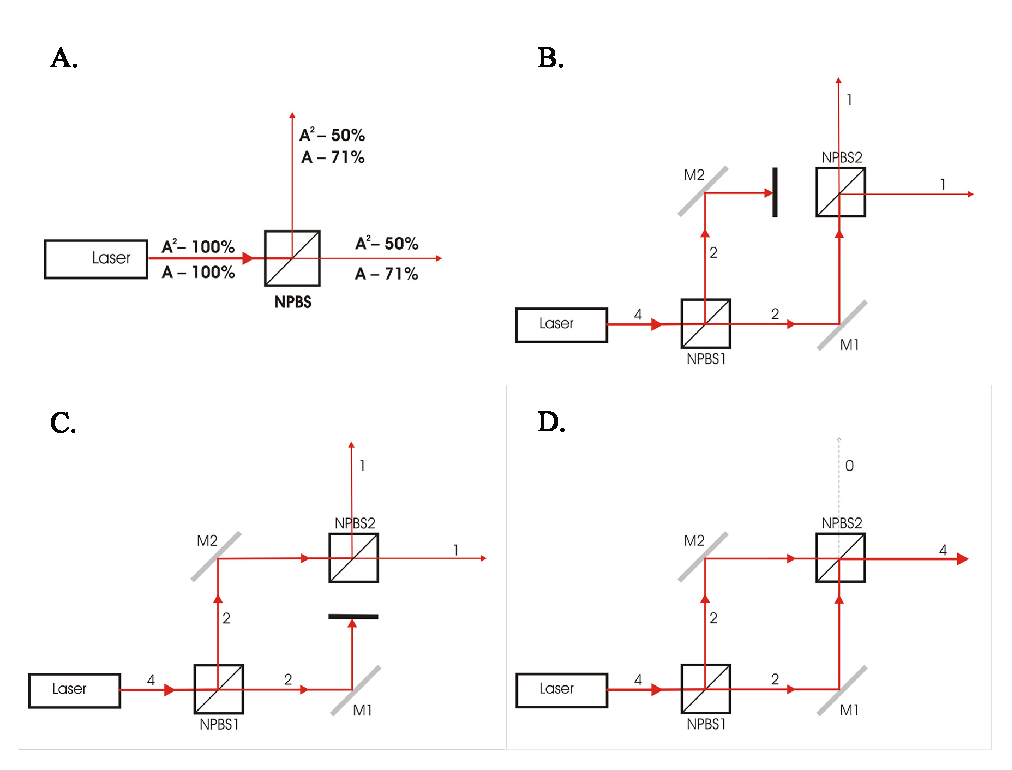}
\caption{\label{fig:mzi} (Color online) \textbf{The amplitude paradox.}
When a laser beam interacts with a beam-splitter (A), it is
observable that the power of the beam is split 50-50. Similarly, if
the transmitted beam (B) or the reflected beam (C) of the first
beam-splitter is allowed to pass through the second beam-splitter of
a Mach-Zehnder Interferometer, the power readings in each channel
drop in half again. Yet, when two beams are superposed in phase in
the same output channel (D), their total energy doubles, as if the
bright channel absorbed all the radiation from the dark channel. If
energy redistribution is denied, then interference must be due to
linear amplitude effects. This means that amplitudes must be split
50-50 at each beam-splitter, dropping by a factor of $2$, rather
than $\sqrt{2}$. Unfortunately, this assumption entails predictions
that contradict the observed energy levels for individual as well as
superposed beams. The only way to derive the correct amplitude
values that work for linear amplitude addition is by assuming that
the reverse version of the same physical process is paradoxically
non-linear.}
\end{figure*}

If this conclusion is accepted as an ontological element, how can it
be justified? One possible strategy would be to assume that wave
crests always obey the rule of linear addition. In other words,
coherence might bring out an essential property of waves, which
could simply be hidden in the case of incoherent wave overlap.
Indeed, wave crests always add up during superposition, but is it a
fact of Nature that this addition is always linear with respect to
\emph{amplitudes}? Consider the case of a collimated laser beam. If
it crosses a 50-50 beam-splitter (Fig.~\ref{fig:mzi}A), we should assume
that the amplitude is split in half, as the beam is divided in two
output projections.
Yet, the proportionality of irradiance to the square of the
amplitude implies that the power of each output beam should drop to
25\%  of the input beam. This corresponds to an expected net energy
loss of 50\%, compared to the input amount. Of course, the actual
observation is that \emph{irradiance} is split 50-50 by the
beam-splitter, without unusual losses. This means that the actual
amplitude of each output must be about equal to 71\% of the input
value, when a beam is split in half. Conversely, when the
beam-splitter is removed from the path of the laser beam, the
amplitude returns to its original value. The energy of the input
beam is not amplified. Yet, this case is physically similar to the
one in which two coherent components overlap. If the amplitudes were
added before squaring, as suggested by rule (\ref{eq1}), quoted
above, the total energy should double. Again, there is a conflict
between interpretive expectations and reality. To be sure, this is
not an attack on the quantitative parameters of superposition. If
the amplitude of a beam is doubled, of course it must have a
quadruple amount of energy. The difficulty is to account for the
reverse relationship: a beam must quadruple its energy before its
amplitude can double. This holds for single beams, whose power can
be modulated, and it must also hold for superposed beams.

The implications of this problem can be illustrated with the
following example. At the first beam-splitter in a Mach-Zehnder
Interferometer (MZI) the input beam is split 50-50. Yet, at the
second beam-splitter the two halves become superposed and
co-propagate. If only one path is open alone, it can be seen that
each beam is split 50-50 again (Fig.~\ref{fig:mzi}B-C). If both paths
are open at the same time, we see the appearance of interference
(Fig.~\ref{fig:mzi}D). The net irradiance in each output path is
determined by the measurable irradiance of individual components, in
accordance with the interference equation:
\begin{equation}
I_{t}=I_{1}+I_{2}+2\sqrt{I_{1}I_{2}}\cos\theta~. \label{eq7}
\end{equation}
In one output, the two components are in phase. They display
constructive interference, and their total energy doubles (1+1=4,
not 2). In the other one, they are out of phase and no radiation is
detectable (1+1=0, not 2). The bright channel contains all the
energy that enters the interferometer, as if it absorbed the
radiation from the dark channel. Yet, the principle of
non-interaction of waves forces us to assume that only the
\emph{detectable} energy is canceled in one output, because the
amplitudes of the two components point in opposite directions.
Similarly, the \emph{detectable} amount of radiation seems to have
doubled in the bright channel, because the amplitudes point in the
same direction and they must be added before squaring in this case.
This means that superposition is not physically effective at the
level of irradiance, as suggested by equation (\ref{eq7}), but
rather at the level of the amplitudes themselves, as suggested by
the equivalent equation (\ref{eq1}). Yet, when we define the
amplitudes as the physical agent of linear superposition, the whole
beam dynamics has to be interpreted in the same terms. For example,
if the starting amplitude is equal to 4 conventional units, then
each component should emerge with 2 units after NPBS1. This implies
that the detectable energy after NPBS1 should drop to 25\%, contrary
to the actual observation of 50\%. Furthermore, the amplitude should
be split again in half after NPBS2, falling to a value of 1 unit, in
which case the observed irradiance should become equal to about 6\%
of the input beam if every path is open alone (by blocking the other
path), contrary to the observed value of 25\%. Yet, when we
calculate the total output energy of superposed components, even by
adding the amplitudes before squaring, we get only half of the
actually observed energy. With or without superposition, it is
impossible to get correct predictions about observable beam
properties, if we assume that amplitudes can really add up and
subtract like vectors. Thus, we get the paradoxical situation in
which we have to assume that amplitudes do not undergo vector
addition, in order to derive correct values that can be used for
vector addition.

This example brings out the crucial ontological difference between
the two main approaches to wave interference. The principle of
interference treats energy as indestructible, with the necessary
implication that it is always redistributed, one way or another. In
the case of incoherent overlap, redistribution is assumed to be so
fast and random that only average effects are detectable in normal
conditions. In the case of coherent overlap, redistribution is
stable over time and follows a strict rule. As a result, we get the
macroscopic \emph{appearance} that amplitudes add up like vectors.
In contrast, the principle of non-interference denies the reality of
energy redistribution. Amplitude addition is assumed to be the
primary physical process, even at the microscopic level. The special
symmetry of this process results in equal amounts of energy creation
and destruction. Therefore, we get the \emph{appearance} of energy
conservation, when total amounts are taken into account. In the
first case, it is energy that propagates, and the net impact of this
process at every point is manifest in the form of amplitudes. In the
second case, it is the amplitudes that propagate as independent
physical entities, and their local net effects are imperfectly
measured in the form of energy. A neutral way to sum this up is that
both phenomena -- amplitude summation and energy redistribution --
are ``appearances''. We have to treat both interpretations as
``stories''. Though, one of these stories is compatible with the
principles of classical mechanics while the other is not. The
principle of non-interference also entails conceptual
inconsistencies, as shown above, because amplitudes ``refuse'' to
display linear manifestations, except in the cases where avenues for
energy redistribution happen to be present. More importantly, the
appearance of energy redistribution cannot be avoided. It can only
be dismissed as unphysical, without the ability to substantiate this
claim. As it is known, the output beams of the MZI remain dark or
bright indefinitely.

\subsection{Energy redistribution is permanent}

A possible objection to the preceding conclusion might be that
beam-splitter interferometers are not appropriate for this
discussion. For all we know, reflecting surfaces or other components
of these devices might really induce energy redistribution. If so,
the non-interaction principle should not be employed for their
analysis. Instead, free-space interference alone should be
interpreted as an example of overlap without interference. The
implied expectation is that energy redistribution can be ruled out
with certainty in this context.

\begin{figure}
\includegraphics{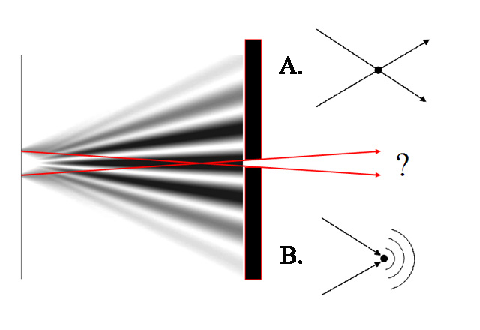}
\caption{\label{fig:fringe} \textbf{Two models of fringe structure.} In
a typical double-slit experiment, superposed wave-fronts do not
separate. The two projections overlap forever like two equal angles
with parallel edges. Still, many different rays can be imagined to
intersect along the central axis of symmetry and it seems that a
perfect lens should able to separate them. If so, the central slice
of a single fringe could be isolated with a slit, in order to test
if its apparent energy is truly illusory. The hidden question is:
what happens when two rays of light from different directions
overlap on a single point? Do they pass through each other
unperturbed and get scattered by edges of the slit (A), or do they
produce indistinguishable spherical Huygens wavelets (B)? In either
case, the two components cannot be separated in actual experiments,
as suggested by the Rayleigh criterion (see below).}
\end{figure}

In many discussions, free-space interference is invoked in reference
to beams that cross each other and then separate completely. Yet,
the interference volume is not always longitudinally finite. The
natural diffraction of radiation and the small angle between the
beams (required for visible fringes) often result in light-cones
that are perpetually superposed (Fig.~\ref{fig:fringe}).
 Nevertheless, the
overlapping projections can be separated with optical devices. This
quasi-transient nature of free-space overlap can be used to suggest
that energy redistribution does not have to be invoked in this case.
Fringes are, of course, detectable, but only upon interaction with a
detector. Conceivably, the latter can be described as an array of
oscillators that absorb and re-emit light. This emission would be
impossible when excitatory radiation components are out of phase,
but it should somehow become amplified by resonance during
constructive interference. For the sake of consistency, one might
even assume that detectors redistribute energy during these
interactions from the areas with destructive interference to the
areas with constructive interference, whereas the energy of incident
radiation is evenly distributed in cross-section. In short, the
appearance of fringes might be an effect of the interaction between
radiation components and detectors \cite{chan06}, rather than
between the propagating radiation components alone. This idea is so
simple that one has to wonder: why are people still debating it? Why
don't they just test it? As it is known, interference patterns can
become arbitrarily large at sufficient distances from the source. If
a fringe is 10 meters wide, then a slit that is 1 meter wide could
isolate the central part of a bright fringe, or that of a dark
fringe. Given the huge ratio of the slit size to the wave-length of
visible light, diffraction can be presumably ignored. Thus, by
aiming a telescope at the median point of such a slit, it should be
possible to separate the fringe slice into two components. According
to the foregoing assumptions, it should be observable that dark
fringes and bright fringes actually contain the same amount of
energy \cite{roy06}. In other words, the bright fringes should
separate into constituent components without extra energy, while the
dark fringes should reveal their hidden energy, without any
possibility of mutual trade-off between different regions in the
interference volume.

\begin{figure*}
\includegraphics{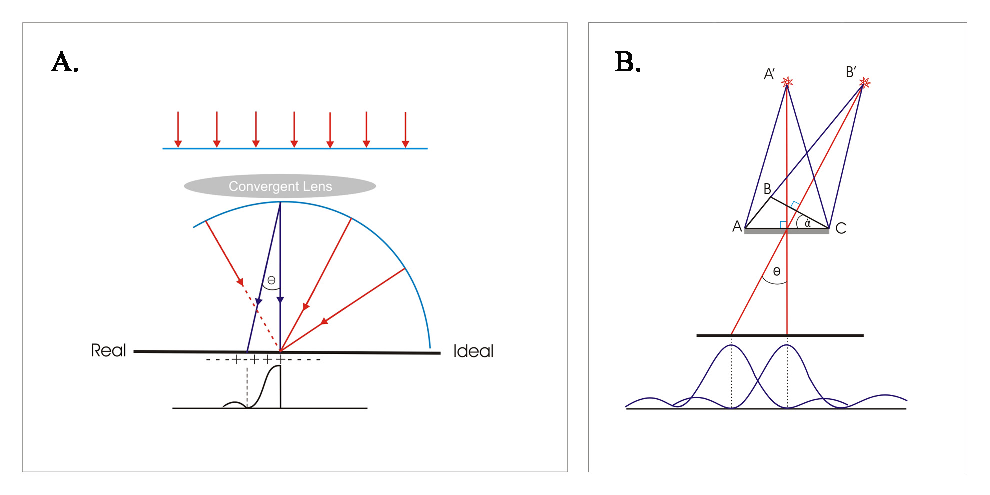}
\caption{\label{fig:rayleigh} (Color online) \textbf{Illustration of the
Rayleigh criterion for resolving diffraction limited spots.} A)
According to Huygens' Principle, a point-like source of radiation
must necessarily produce a spherical front of constructive
interference. A finite circular section of a spherical front can
only be produced by an emitter with real extension, shaped like a
disk with rings. Due to the inbuilt symmetry of this model,
convergent projections play back the same process in reverse. A
negative spherical front of constructive interference converges onto
a single point, as shown in the right side of the drawing. Finite
lenses can only capture a circular section of this pattern. Without
contributing wavelets or rigid boundaries at the edges, the wave
front ``opens up'' (left side of the drawing). The result is a
diffraction (Airy) pattern that looks like a bright disk with faint
rings. The angular size of the bright spot, from the central point
to the edge of the first ring of destructive interference, is
determined by the relationship $sin{\theta}=1.22\lambda/d$, where
$d$ is the diameter of the lens. B) Two point-like sources of
radiation, separated by a very small angle, produce overlapping Airy
patterns that may be impossible to isolate. The Rayleigh criterion
stipulates that such sources are going to be minimally resolved if
their angular separation is equal to the angular size $\theta$ of
the diffraction limited spot from drawing A. The criterion ensures
that the peaks of the two Airy patterns fall in each other's first
ring of destructive interference. The drawing B illustrates that the
angle $\theta$ is also equal to the angle $\alpha$ between two plane
waves from the two sources. If these waves are in phase at the right
edge of a lens, they must be out of phase by $1.22\lambda$ at the
left edge. Accordingly, two wave-fronts cannot be resolved if they
have a range of phase-differences smaller than $1.22\lambda$ across
the surface of a lens. For a uniformly illuminated slit, a range of
$1.00\lambda$ is sufficient. Due to the symmetry of the Huygens
model, it follows that radiation from two sources, passing through a
single slit, must be subjected to the same restriction. Two coherent
laser beams, no matter how far from each other at the source, will
not be able to produce separable projections through a slit of any
size, unless the width of their interference fringes is smaller than
the width of the slit. Even if such a projection were passed through
a lens with infinite diameter, the necessary range of phase
differences for beam separation cannot be physically available in a
small fraction of a single fringe.}
\end{figure*}

Several variations of this proposal have been advanced recently
\cite{peng09, pras11, amb11}, but no empirical verification has been
reported. To the best of our knowledge, this is not for lack of
trying. In actual experiments, bright fringes cannot be split into
components with less total energy, and dark fringes remain dark. Why
would that be the case? If a lens with a diameter of 1cm can
separate an interference volume into its components (close to the
source), why would it be so difficult to separate a 1m slice of the
same volume (far from the source)? According to the principle of
non-interaction, the two cases should be qualitatively similar,
because component wave-packets are supposed to just pass through
each other in the same way in both situations. If anything,
scattering effects should be smaller in the case of wide
slits/apertures. Could it be that something has been overlooked in
this analysis? The answer can be found by reviewing some basic facts
in the theory of optical resolution. As it is well known, a
self-luminous point cannot be imaged precisely in the far field with
a lens that has a finite diameter (see, for example \cite{hecht}, or
any other optics textbook). A point is always imaged as an Airy
pattern, whose spread is directly proportional to the wavelength and
inversely proportional to the diameter of the lens. The angular size
of the central bright disc of the Airy pattern
(Fig.~\ref{fig:rayleigh}A)
 is given by:
\begin{equation}
\sin\theta=\frac{1.22\lambda}{d}~. \label{eq8}
\end{equation}
For a given source of light with fixed wavelength, it is only the
size of the lens/aperture ($d$) that can be changed to improve the
resolution. When the diameter of the lens is comparable to the
wavelength of light, it becomes useless because it cannot resolve
any point. A practical consequence of this feature is that radio
telescopes need to be very large compared to optical telescopes that
work with short wavelengths. Yet, there is an important difference
between the absolute limits of resolution of a lens and the
constraint of resolving small objects. In most practical situations,
the challenge is not to resolve just a single point. Instead, it is
to resolve two points that are very close to each other. Lens
diameter matters greatly here too, of course. Bigger lenses produce
smaller Airy patterns, increasing the resolution power. Yet, a very
large lens can still fail to resolve two points if their angular
separation is small enough. According to the Rayleigh criterion
\cite{serway}, two points are ``just resolved'' when their angular
separation, as seen from the central point of the lens, is equal to
the angular size of the Airy disc (Fig.~\ref{fig:rayleigh}B), such that
the central points of each projection overlap with the first dark
rings of the other:
\begin{equation}
\sin\alpha=\sin\theta=\frac{1.22\lambda}{d}~. \label{eq9}
\end{equation}
This arrangement has a remarkable geometrical property. As shown in
Figure~\ref{fig:rayleigh}B, the angular separation of the two sources is
equal to the angle of the tilt of the lens between alignment A, when
it is normal to the first source, and alignment B, when it is normal
to the second source. This is equivalent to the angle between two
co-incident plane waves, originating at the same sources. If these
waves are in phase at one edge of the lens, they will have a maximal
phase delay at the other. When the angular separation between the
sources is at the critical value, as specified by equation
(\ref{eq9}), this phase delay is equal to 1.22 wavelengths. In other
words, for any wavelength, and for any lens diameter, two sources
are only going to be resolved if they have a range of phase delays
in excess of 1.22 wavelengths across the diameter of the lens. As it
is well-known, the width of an interference fringe corresponds to
the cross-segment in which the range of phase differences between
two coherent components is equal to one wavelength. This parameter
is fixed, regardless of the size of the fringes. Consequently, it
does not matter how far from the sources one places a screen and how
big the fringes are. Even with the largest lens in the world, a
narrow slice of a single fringe can never be separated into its
components, because it can only contain a negligible range of phase
delays. The physical significance of this conclusion can be
understood by recalling that wavelet phase delays determine the net
direction of a wavefront in Huygens' model. If two wavefronts
display negligible variation in their phase differences, they must
co-propagate and remain forever indistinguishable, even if they do
not interfere.

When linear superposition is analyzed with a particle model such as
ray tracing, it is convenient to describe the beams ``as if'' they
pass through each other at every point of intersection.
Unfortunately, this assumption contradicts the Rayleigh criterion. A
possible loophole might be to assume that scattering effects at the
edges of a slit or a lens are hiding the process of rectilinear
propagation. The trade-off is to revert to a wave model, because
particles do not cause diffraction fringes. Even so, the role of
edge scattering becomes less and less compelling as the size of the
fringe is allowed to increase with propagation. In order to maintain
consistency with known empirical facts, the principle of
non-interference can only be applied on the basis of a wave model
that is consistent with Huygens' Principle. In this case, the
explanation becomes that different beams obey Huygens' Principle
independently. They are still presumed to be real in the dark
fringes. It just so happens that human observers cannot isolate them
for technical reasons. The problem with this story is that Huygens'
Principle describes the direction of a wave as a collective process.
It also sums up the contributions from different directions onto a
point of incidence in a single spherical wavelet. This means that
input waves can only pass through each other if indistinguishable
components of single spherical wavelets ``remember'' where they came
from and separate accordingly in subsequent interactions. Yet,
somehow they still cannot regain their input direction unless dark
fringes are allowed to merge with bright fringes.

As a reminder, the starting point of this section was an attempt to
show that non-interference is experimentally verifiable. The price
was to restrain the validity of this concept to free space
interference, excluding all the instances of beam-splitter
interference, even though the same equations are required in both
contexts. It is apparent now that free space interference is also
too wide a concept for this task. In most cases, when point-like
sources radiate light, the outcome is an infinite volume of
superposition in which dark fringes persist like infinite
projections, without ever mixing with bright fringes, just like in
the case of beam-splitter interference. States with new levels of
energy do not ``naturally'' separate back into original components.
Moreover, it is even impossible to separate them with optical
devices, in another example of similarity with collinear
beam-splitter superposition. Thus, energy redistribution during
infinite interference in free space is permanent. If the
non-interaction principle is used to interpret this observation,
then it is necessary to invoke non-locality.

All of the above notwithstanding, the appeal of non-interference
comes from the special class of situations where collimated or
focused beams intersect with a finite volume of overlap. In this
case, fringes are observable in superposition, but the beams are
also able to separate without intervening lenses. Can it not be that
the concept of energy redistribution is dispensable in this case?
Surprisingly, the answer is negative again. If portions of the dark
fringes or the bright fringes are isolated with slits, it becomes
impossible to observe beam separation, just like in the case of
free-space interference with infinite interference volumes, as
discussed above. The only exception to this rule happens when
several bright fringes are open at the same time, with narrow
obstacles at the center of dark fringes. In this case, as shown
recently by Afshar \cite{afs07}, beam separation becomes visible,
but energy redistribution is a necessary implication because the
obstacles do not appear to remove the expected amounts of energy
from the superposed beams and do not produce diffraction fringes.
Either radiation is flowing around the dark fringes (assuming local
energy redistribution in the interference volume), or the apparent
beam separation is the outcome of interference between the elements
that were diffracted by the slit edges \cite{steu07} (resulting in
non-local energy redistribution beyond the original interference
volume) \footnote{This example will be explained in greater detail
in the next chapter.}. Consequently, it is not possible to prove the
absence of energy redistribution, even when the superposed beams
separate without intervening optical devices. Perhaps, this can be
explained away as a limitation of our means of observation, by
assuming that measurements are always intrusive. Nevertheless, the
point remains that the principle of non-interaction does not entail
verifiable new physics.

A useful lesson from this discussion is that the observable details
of interference are always similar when the quantitative details are
similar. Whether we analyze the superposition of radiation in
beam-splitters or in free space, the relevant equations are the same
and the subtle physical features are also the same. A region with
perfect destructive interference is mathematically and physically
equivalent to a region without any radiation, no matter what
operations we perform over it. In the context of our discussion,
this means that the appearance of energy redistribution cannot be
avoided in any context. Regions with missing or excessive energy can
only revert to their baseline levels if they have the opportunity to
mix with each other at later stages of propagation. As a corollary,
the principle of non-interaction cannot make predictions that are
distinguishable from models that assume real energy redistribution.
The only difference is that non-interaction entails non-locality in
this process.

In conclusion, modern classical physics interprets wave interactions
with concepts that sometimes violate the spirit and the letter of
Newtonian mechanics, despite the existence of alternative models. It
may seem harmless to assume that waves are transparent to each
other. Yet, this entails non-local energy redistribution, transfer
of kinetic energy without physical motion, as well as simultaneous
motion in multiple directions for single macroscopic bodies. No less
importantly, conceptual inconsistencies follow from the differential
treatment of single beams and superposed beams. How can it be that a
proven classical model violates the principles of classical physics?
We suspect that many models in contemporary science are designed to
interpret successful equations, rather than actual physical
processes. The purpose of formal analysis is to simplify
calculations, rather than to capture the whole complexity of the
Universe. Sometimes this entails replacing cosmic bodies with
point-like objects; sometimes it involves crossing out real forces
that balance each other out; and sometimes it leads to the analysis
of standing oscillations in terms of running waves that pass through
each other unperturbed in opposite directions. What is important to
acknowledge is that the same equations can also be interpreted with
stories that satisfy the principles of classical physics.

\section{Superposition: the test of parsimony}

Classical interpretations possess a higher level of intuitive appeal
than non-classical alternatives, but this does not prove they are
ontologically superior. For all we know, human psychological
predispositions could be accidental, even if shared across cultures.
Instead, the competition between two equivalent theories can be more
conclusively resolved on the basis of the principle of parsimony,
otherwise known as Ockham's razor. If one theory explains a greater
number of phenomena with fewer assumptions, it is a more compelling
description of reality. This principle is \emph{also} based on an
intuition: it makes little sense to explain something ``the hard
way'' (\emph{e.g.} one independent theory for each phenomenon), if
it is possible to do it ``the easy way'' (\emph{e.g.} one common
theory for all the relevant phenomena). Though, it has the authority
of a long-standing standard of last resort in science, where
hallmark discoveries are commonly associated with insights that
reduce complex appearances to simple common mechanisms. Accordingly,
it is not enough to conclude that one approach to the nature of wave
interference is classical, while the other is not. If the principle
of non-interference was found to be more parsimonious, it would be
reasonable to treat it as ontologically preferable despite its
non-classical features, as long as it did not entail any
contradiction with observable phenomena. In any case, this is how
modern physics is generally interpreted.

The purpose of this chapter is to determine: which of the two
interpretations of linear superposition can survive Ockham's razor?
If waves are assumed to pass through each other unperturbed,
associated quantitative tools are often simpler. The question is: do
we get a simpler ontology? Do we acquire the ability to explain more
phenomena with a common mechanism, when compared to the assumption
that wave interference is real? This problem is going to be tackled
by reviewing the properties of superposed optical beams
distinguishable by polarization, frequency, and direction of
propagation. It will be shown that superposition results in
observable net states that can be reproduced with single beams as
well. The physical properties of these two types of phenomena are
identical, quantitatively as well as experimentally. Yet, the
principle of non-interference requires different underlying
mechanisms for each of them. In contrast, the assumption of real
interference requires a single straightforward mechanism. Therefore,
the non-classical interpretation is also the one that fails the test
of parsimony.

\subsection{Beams with different polarization}

\begin{figure}
\includegraphics{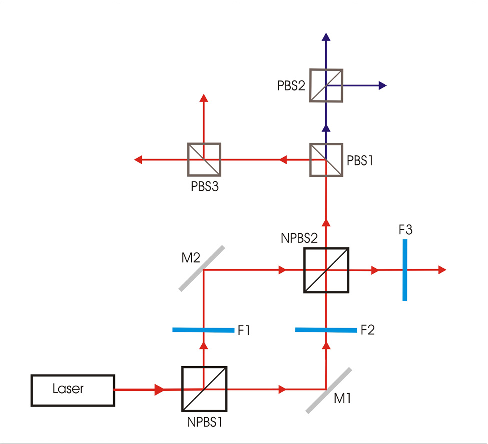}
\caption{\label{fig:mzi04} (Color online) \textbf{Beam-splitter
interference between collinear polarized beams.} Two non-polarizing
beam-splitters (NPBS1 and NPBS2) and two mirrors (M1 and M2) form a
Mach-Zehnder interferometer with equal paths. Radiation components
from each path co-propagate at the output. In one channel (towards
filter F3) superposed components are in phase. In the other channel
(towards PBS1) they are out of phase. Polarizing filter F1 has a
horizontal axis of polarization. Filter F2 is vertically polarized.
Filter F3 is diagonal. When input components are in phase, they
behave like an output beam with diagonal axis of polarization, with
or without filter F3. When the components are out of phase, they
behave like a beam with linear polarization, orthogonal to the
diagonal axis. Polarizing beam-splitter PBS1 transmits diagonal
polarization and reflects anti-diagonal components. PBS2 is aligned
to split input light into vertical and horizontal components. No
radiation is observed to arrive in or out of PBS2. PBS3 is also
aligned to split input radiation into vertical and horizontal
components. It appears to receive all the light that enters PBS1
(after reflection) and splits it 50-50.}
\end{figure}

The most convenient tool for studying the interaction of polarized
beams, in our experience, is the Mach-Zehnder interferometer with
collinear output beams in each channel. Consider the set-up shown in
Fig.~\ref{fig:mzi04},
in which linear polarizing filters are installed in each path, as
well as in the output channels. Assume that the interferometer is
properly aligned, such that the two component beams are in phase in
one output channel (towards the filter F3, and out of phase in the
other. If the axis of one polarizer (F1) is parallel to the
horizontal plane, and the other one (F2) is parallel to the vertical
plane, the filtered beam polarizations are orthogonal to each other.
What happens when these beams arrive at the analyzer F3? We apply
Malus' Law to predict the amount of light that will be able to pass
through. As a reminder, this law states that the proportion of
transmitted light must be equal to the cosine squared of the angle
between the plane of polarization of the beam ($\alpha$) and the
axis of the analyzer ($\gamma$):
\begin{equation}
I_{out}=I_{in}\cos^{2}(\alpha-\gamma)~. \label{eq10}
\end{equation}

When $(\alpha-\gamma)=45^{\circ}$, the proportion of transmitted
radiation is 50\% (ignoring the unavoidable losses at the
intervening optical surfaces). Indeed, only half of each beam is
transmitted, when they are open one a time. The outcome is the same
if the beams are filtered in the diagonal plane, or the
anti-diagonal plane \footnote{``Anti-diagonal'' represents the
direction that is orthogonal to the diagonal plane.}, because the
absolute value of $(\alpha-\gamma)$ is the same in both situations.
However, things are very different when both beams are open at the
same time. Two coherent linear states of polarization produce a net
state that is also linearly polarized, in the median plane, if the
superposed components are in phase. In this case, the net state
corresponds to the diagonal plane. This means that all the radiation
must pass through the filter F3, if the latter is aligned in the
diagonal plane, and all of it must be blocked if the filter is
aligned in the anti-diagonal plane, in accordance with Malus' Law.
This expectation is confirmed by experimental observations. For all
intents and purposes, the sum of the two output beams behaves as if
it was a single beam with diagonal polarization. At the same time,
the net state in the other channel of the interferometer (where the
components are out of phase) behaves as if it was polarized in the
anti-diagonal plane, as confirmed by the amount of light reflected
by PBS1.

If we assume that interference is real, then component states are
not physically relevant. This explains why the net state is able to
pass entirely through the diagonally oriented analyzer F3: no
radiation is being reflected. In contrast, if interference is
assumed to be an illusion, then beam components must behave in the
same way, whether present alone or in superposition. During the
interaction with F3, only half of the horizontal component must be
assumed to pass through, as well as half of the vertical component,
but the resulting components with parallel polarization must be
assumed to produce the appearance of interference. If the irradiance
of beam 1 in the relevant channel is designated as $I_{1}$, the
amount of light that passes a polarizing filter is determined by
$I_{1}\cos^{2}(\alpha-\gamma)$. If the irradiance of beam 2 is
designated as $I_{2}$, then the amount of this beam that passes the
same filter is given by $I_{2}\cos^{2}(\beta-\gamma)$. Note that
both components are polarized in the plane of the analyzing filter
at the output. If these values are plugged into the interference
equation for beams with parallel planes of polarization:
\begin{equation}
I_{t}=I_{1}+I_{2}+2\sqrt{I_{1}I_{2}}\cos\theta \label{eq11}
\end{equation}
the amount of light transmitted by F3 is given by:
\begin{eqnarray}
I_{t}=I_{1}&& \cos^{2}(\alpha-\gamma)+I_{2}\cos^{2}(\beta-\gamma) \nonumber\\
&&+~2\sqrt{I_{1}I_{2}}\cos(\alpha-\gamma)\cos(\beta-\gamma)\cos\theta
\label{eq12}
\end{eqnarray}
where $\theta$ is the phase difference between the component beams
in superposition. This expression can also be derived by expanding
equation (\ref{eq10}) for the net state, where the amplitude of the
net state is represented as the vector sum of the two component
amplitudes. Hence, the quantitative predictions are necessarily
equivalent in either scenario, with or without real interference. In
the specific example described above, with two orthogonally
polarized beams and a diagonal analyzer, the long expression
(\ref{eq12}) simplifies to:
\begin{equation}
I_{t}=\frac{I_{1}}{2}+\frac{I_{2}}{2}+\sqrt{I_{1}I_{2}}\cos\theta~.
\label{eq13}
\end{equation}
Note that Malus' Law predicts the amount of light that is blocked or
reflected by an analyzer as proportional to the \emph{sine} squared
of the original amount. This implies that the amount of reflected
light for two beams in linear superposition must be:
\begin{eqnarray}
I_{t}=I_{1}&&\sin^{2}(\alpha-\gamma)+I_{2}\sin^{2}(\beta-\gamma) \nonumber\\
&&+~2\sqrt{I_{1}I_{2}}\sin(\alpha-\gamma)\sin(\beta-\gamma)\cos\theta~.
\label{eq14}
\end{eqnarray}
This can be verified empirically by rotating the analyzer F3, or by
replacing it with a polarizing beam-splitter. Again, for the
specific example considered above, expression (\ref{eq14})
simplifies to:
\begin{equation}
I_{t}=\frac{I_{1}}{2}+\frac{I_{2}}{2}-\sqrt{I_{1}I_{2}}\cos\theta~.
\label{eq15}
\end{equation}
Equations (\ref{eq13}) and (\ref{eq15}) exhaust all the energy that
is present in the channel under consideration, and they are
empirically accurate. If they are taken into account together, it is
hard not to notice that the same amount is subtracted from one
expression and added to the other. This relationship also holds for
the outputs of PBS1 in the second channel of the interferometer. In
short, we can say that the net state is an illusion, because
component interference results in just the right amount of added or
missing energy to mimic its behavior. According to the principle of
non-interference, each input beam is split by PBS1 in the same way,
whether one of them or both are open at any point in time. It just
so happens that the net state in the transmitted path cancels out
the same amount of energy that is revealed in surplus in the
reflected path, when the components are simultaneously present. In
the considered example, where $\cos\theta=-1$, we get one dark
channel and one bright channel that seems to acquire all the energy.
Though, we should assume that this apparent transfer is
non-physical, because the two components are presumed to continue
their propagation unperturbed in each channel. Also, as discussed in
section II, we should pretend that \emph{amplitudes} are split 50-50
in order to explain this process without interference, even though
the equations (\ref{eq11})~--~(\ref{eq15}) work at the level of
\emph{irradiance}.

It might seem that the non-interaction principle is verifiable in
this context. In the reflected path of PBS1 (Fig.~\ref{fig:mzi04}) with
the fast axis in the diagonal plane we have the appearance that the
energy has doubled, like in any other bright fringe. In the
transmitted path we have the appearance that energy has vanished,
like in any other dark fringe. Yet, either output beam can be split
again with polarizing beam-splitters, in an attempt to reveal the
hidden components. For example, the dark projection can be split
again with PBS2, aligned with the fast axis in the vertical plane.
That should make the energy visible again, effectively extracting
energy from a dark fringe. Similarly, by splitting the bright beam
into vertical and horizontal sub-components, it should appear to
lose half of its energy. Though, if that was the case, then a
violation of energy conservation would follow when the dark fringe
was split while the bright fringe was not. In actuality, the dark
channel remains dark forever, and the bright fringe does not loose
energy spontaneously. This can be explained by taking into account
the details of the interaction between the input beams and the
polarizing beam-splitter. Prior to reaching PBS1, the vertical and
the horizontal beams are assumed to be distinguishable. If the PBS1
were rotated with the fast axis in the horizontal plane, the mixture
would separate into its components, and we would assume to know
where each output comes from. Vertically polarized light must come
from F2 and horizontally polarized light must come from F1. (In
quantum mechanics, the equivalent of this would be path knowledge).
In contrast, the diagonal components from each beam are
indistinguishable behind PBS1 (if it is rotated back to transmit
diagonally polarized light). The same holds for the anti-diagonal
components in the reflected path. Ergo, PBS2 is still assumed to
receive 50\% of the beam energy when the channel is dark, but the
components with parallel polarization are assumed to cancel out even
after being split, producing the appearance of zero energy at each
output. Similarly, PBS3 can split the bright channel into two
components, one vertical and horizontal. Yet, these components do
not revert to lower levels of energy, because they result from
indistinguishable diagonal components form both input beams. In
other words, we must assume that our measurements of irradiance are
wrong, because the energy only appears to be doubled due to
interference. Yet, it will stay wrong forever, because the two
components cannot be made distinguishable again.

To sum up, the principle of non-interference implies that we can
never know what we measure. A linearly polarized beam could be a
true single beam, whose measurable energy corresponds to its actual
content. Though, it could also be a coherent mixture of two
indistinguishable beams with half the ``true'' total energy. No
physical marker is available for distinguishing these states.
Indeed, a true linearly polarized beam can also be described as if
it was made of indistinguishable components. The only difference
between these states is found in their history. One beam is
traceable directly to it source, while the other has passed through
several optical devices such as to produce a twin beam with zero
detectable energy (and half the ``true'' energy). On the other hand,
it is possible to avoid the need for such distinctions, if the
interference principle is assumed to be valid. In this case, the net
state of two overlapping beams (or more) must always be treated as
real. Simultaneous oscillations in different directions at the same
point are not possible in this context. So, when the two
orthogonally polarized components are mixed at NPBS2, they must
suffer the necessary physical transformation, in order to become a
single-state beam with linear polarization in the diagonal plane. As
such, the net state is reflected unchanged by the PBS1, without
anything available for transmission. This state cannot be assumed to
contain any path knowledge, because the first filtration is
identical to the second. Moreover, there is no need to assume the
reality of undetectable dark beams with hidden energy. If two beams
appear to be identical, then they must be assumed to be physically
identical.

\begin{figure*}
\includegraphics{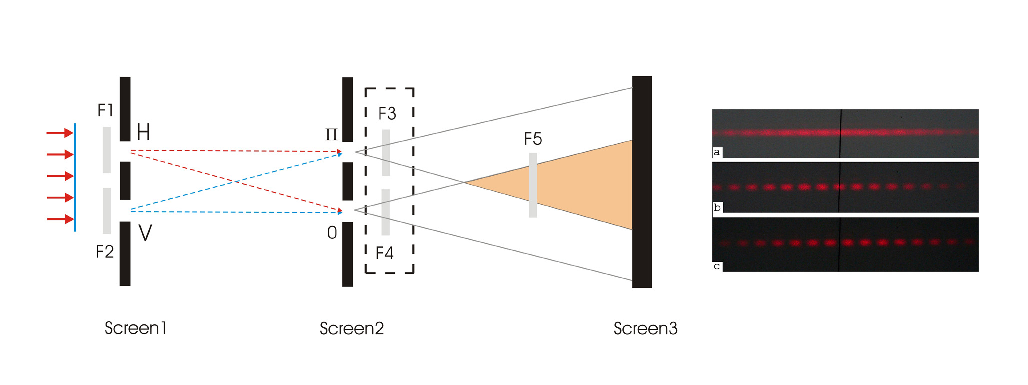}
\caption{\label{fig:screen} (Color online) \textbf{Free-space
interference between polarized beams.} This is a superposition of
two double-slit experiments. The two slits at Screen1 are covered
with orthogonal linear polarizers. The upper slit emits radiation
with horizontal polarization. The lower slit projection is
vertically polarized. At Screen2, the lower slit is centered on the
point where components from V and H have zero path difference. The
center of the upper slit corresponds to a point with a path
difference of $\pi$. If the H slit at Screen 1 is open alone, a
double slit interference pattern can be seen at Screen 3, in the
absence of filters F3, F4 and F5. Similarly, slit V alone produces a
complementary fringe pattern. If both slits are open at the same
time, a continuous pattern is observed (image [a], to the right).
Filter F5 can reveal a set of fringes if it is horizontally aligned
(image [b]), and a set of anti-fringes in the vertical alignment
(image [c]). If looks as if F5 blocks light from the V slit, or H
slit, as the case may be, revealing undisturbed components from each
source. Yet, the projection from the lower slit at Screen 2 is
linearly polarized in the diagonal plane. 100\% of it passes through
the filter F4, which has a diagonal axis of polarization. The upper
slit projects a beam with anti-diagonal polarization. It also
appears to pass entirely through filter F3 (with anti-diagonal axis
of polarization). The presence or absence of filters F3 and F4 has
no qualitative effect on the observable projections at Screen3, when
both slits are open at Screen1.}
\end{figure*}

This conclusion is not just a peculiarity of superposition in the
MZI. A similar situation emerges if a Young interferometer is used
instead. As shown in Fig.~\ref{fig:screen},
a coherent source of light can be blocked by a screen with two
slits, where each slit is covered by a linear polarizer. If one
filter is oriented vertically, while the other is horizontal, the
far-field projection of this double-slit setup appears featureless,
without fringes. Still, if the transverse plane of the projection is
scanned with a narrow slit and a linear polarizer, it will be
observed that some regions are linearly polarized in the diagonal
plane (where vertical and horizontal components are in phase),
others are anti-diagonally polarized (where they are out of phase),
with elliptical states of polarization in-between. This property can
be exploited by placing a second screen with two slits in the path
of the projection. One slit must be carefully positioned in a region
where the output is diagonally polarized, while the other should
fall on an adjacent region with anti-diagonal polarization. This
time, the slits are not covered with any polarizing filters, but it
can be confirmed with additional measurements that output light is
linearly polarized, as described, in two orthogonal planes (diagonal
and anti-diagonal). The far field projection of the two slits from
the second screen will be featureless again, but the whole
projection can be filtered with a linear analyzer (F5), revealing
fringes in the horizontal plane. Similarly, by rotating the analyzer
to the vertical plane, a set of anti-fringes will be discovered.
What is the physical process behind this observation? Assuming
non-interference, we must take it for granted that diagonal and
anti-diagonal polarizations behind the second screen slits are
illusory. We must assume that horizontally polarized light from the
\textbf{H} slit of the first screen has passed through both slits,
producing the appearance of bright fringes. Hence, all the light
that is seen in the presence of the horizontal filter must come from
that slit. The filter has blocked all the light from the \textbf{V}
slit. On the other hand, when the filter is rotated to the vertical
plane, the light from the \textbf{H} slit is blocked and we see
fringes produced only by light from the \textbf{V} slit of the first
screen. Here is the twist: the slits on the second screen can be
covered with linear polarizers as well. If a diagonally polarizing
filter is placed over the slit that emits diagonally polarized
light, the projection remains unchanged, as far as physical
observations are concerned. Similarly, the slit with anti-diagonally
polarized radiation can be covered with a filter that is oriented in
the same plane. After this modification, the far field projection
suffers no consequence, and the same types of fringes will be
observed with the linear filter used for analysis. Yet, this time we
must assume that we are dealing with indistinguishable half-beams
with illusory extra energy, like in the MZI set-up described above.
Hence, we must describe the same fringes as a result of
superposition of diagonal and anti-diagonal polarizations, rather
than original vertical and horizontal polarizations. In the absence
of filters at the second screen, we are supposed to have path
knowledge. In their presence, the situation is radically different,
even though every subsequent measurement would produce the same
observations. Again, a pure classical interpretation would suggest
that we have the same physical process in both situations, because
the net state is always the only real state. We do not need to know
how the plane of polarization became diagonal or anti-diagonal,
because that does not change its nature. As explained in the
previous chapter, the net state of two superposed beams cannot be
expected to separate into input components behind a slit that is
narrower than an interference fringe. Whether we assume that the net
state is real or illusory, observable measurement outcomes are the
same.

In conclusion, the assumption of non-interference requires a very
complex model of superposition for polarized beams. At every
interaction with measuring devices, we have to allow for the
possibility of unobservable projections with ``real'' energy,
implying that the energy of observable projections is not real. Yet,
the unobservable and the observable states are irreversible,
suggesting that all the real effect of bright fringes come from
unreal energy. More importantly for this discussion, the assumption
of non-interference implies that physical observations (and the
equations that fit them) are indeterminate. Whenever we detect a
beam with linear polarization, we cannot know if it is an ``actual''
single-mode projection, or a mixture of two components with half the
total energy, overlapping in phase. Two different mechanisms are
required for the same type of observations. Moreover, we cannot
determine which is which by doing measurements, even though this is
a classical state. The only way to distinguish the two hypothetical
scenarios is by acquiring the full prior history of the projections,
all the way to the source. Yet, after all the extra work, we cannot
expect a matching compensation, because this interpretation does not
entail any new prediction. The subsequent behavior of the beam in
question is going to be the same in any experimental setting,
regardless of its mechanism of emergence. None of these
complications are found in the alternative scenario, where
interference is assumed to be real.

\subsection{Beams with different frequencies}

\begin{figure}
\includegraphics{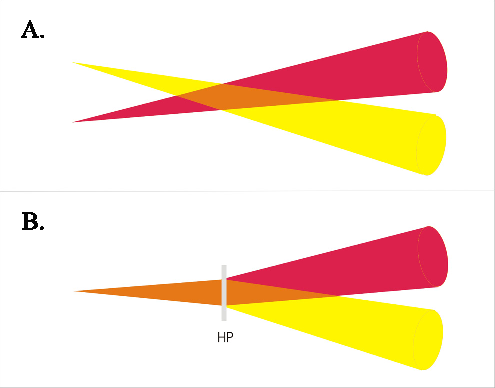}
\caption{\label{fig:freq} (Color online) \textbf{Interference between
beams with different frequency.} A monochromatic red beam intersects
a monochromatic yellow beam. (A) The net state of superposition
behaves like a monochromatic orange beam that is variably amplitude
modulated in the cross-section. Yet, the output beams look identical
to the input beams, as if they just passed through. (B) The same net
state of superposition can be generated by passing a monochromatic
orange beam through a suitable hologram of intersecting red and
yellow beams. The physical role of the holographic plate HP is to
modulate the amplitude of the image-forming component of the
monochromatic input beam. If the left side of the drawing (B) were
covered, it would seem ``obvious'' that a red beam has passed
through a yellow beam. Still, this is impossible, because red and
yellow components are not physically present at the input side. This
\emph{gedanken} experiment is based on real findings, some of which
are referenced in the text.}
\end{figure}

The concept of non-interference appears to be less confusing in the
case of interactions between beams with different frequencies.
Consider the interaction of two monochromatic laser beams with
visible differences in color (for example, red and yellow, as in
Fig.~\ref{fig:freq}A), which intersect at a narrow angle.
Each beam can be traced visually along its path. Of course, colors
are mixed in the volume of overlap, but they are unmistakable before
and after the interaction. The red beam is clearly seen going in and
coming out unchanged (especially if dust particles are present in
the air). The same is true about the yellow beam. How can a beam
come out unchanged and not be the same? Furthermore, one beam can be
analyzed with a detector while the other is blocked intermittently
at the source. The open beam looks the same whether it is present in
the interferometer alone or not. For this reason, it seems
counterintuitive to even suggest that the beams do not go through
each other unperturbed. Yet, the ultimate physical nature of
electromagnetic radiation is still not perfectly understood. For all
we know, it could be based on a process that involves real waves or
is otherwise compatible with the basic principles of classical
mechanics. If so, then we have to allow for the possibility that two
modes of oscillation are not simultaneously possible in the same
point in space, and only the net state of the superposition of two
waves is physically real. In other words, the beams may appear
identical before and after overlap, but it is not an indisputable
fact that the energy in one output comes from a single input. The
beams can be assumed to transform into the net state, which then
evolves into the observable outputs because of the existing
symmetries in the underlying physical interactions. Remarkably, the
quantitative aspects of these two models are equivalent. As shown
above, and also as demonstrated by Dowling and Gea-Banacloche
\cite{dow92}, the two alternative assumptions about interference
entail the same macroscopic observations. Hence the problem is not
that one model makes better predictions than the other. The question
is whether we can increase our \emph{interpretive} power by invoking
non-interference.

The mixture of beams with different frequencies is well understood
theoretically and is commonly analyzed in terms of Fourier
synthesis. The main features of this process are  captured by the
following identity:
\begin{equation}
\cos(\omega_{1}t)+\cos(\omega_{2}t)=2\cos(\omega_{m}t)\cos(\omega_{0}t)
\label{eq16}
\end{equation}
where $\omega_{1}$ and $\omega_{2}$ are the angular frequencies of
two input beams, while $\omega_{m}=(\omega_{1}-\omega_{2})/2$ and
$\omega_{0}=(\omega_{1}+\omega_{2})/2$ \cite{gil07}. In other words,
whenever we mix two beams with different frequencies, the net state
behaves as a new beam with a carrier frequency $\omega_{0}$ that is
equal to the average of the input frequencies. The important feature
of this new state is that its envelope is modulated with a beat
frequency $\omega_{m}$ that is equal to half the difference between
the input frequencies. Accordingly, if one assumes that
non-interference is real, then only the left-hand side of the
equation (\ref{eq16}) is physically significant. In contrast, the
right-hand side of this equation must be assumed to be real if the
interference principle is preferred. The identity relationship
captures the quantitative equivalence of the two models. Is it
possible to tell which model is ontologically accurate, despite
their mathematical equivalence? Recently, Lee and Roychoudhuri
(L\&R) attempted to answer this question with a series of didactic
experiments \cite{lee03}. They reported a remarkable observation. If
a mixture of two different frequencies is analyzed with a wide-band
photo-detector, the recorded pattern is consistent with the features
of the right-hand side of the equation (\ref{eq16}). However, if the
same mixture is detected with a high-resolution Fabry-Perot
spectrometer, the beam is not transmitted at the average frequency
($\omega_{0}$). Instead, only input frequencies ($\omega_{1}$ and
$\omega_{2}$) are able to sustain the necessary resonance for
passing through the cavity. Moreover, the same mixture was unable to
produce fluorescence in a rubidium (Rb) atomic vapor, if the average
frequency ($\omega_{0}$) was centered on a natural Rb transition
line, in contrast to the case when input frequencies ($\omega_{1}$
or $\omega_{2}$) were centered on the same line. In short,
narrow-band detectors could not see the frequency of the net state.
Doesn't this prove that the net state is non-physical? If it was
real, why did it not show up in the relevant experiments? If the
input states are no longer real, why did they produce the observed
resonance?

In order to answer these questions, another group (Gilra \emph{et
al}., \cite{gil07}) performed a similar series of experiments
without mixing two input frequencies. Instead, they used a
monochromatic laser beam whose envelope was modulated with an
acousto-optical modulator. In this manner, they produced an optical
state with the exact properties that would have been possessed by
the average frequency of two different beams. Hence, they created a
beam that corresponded directly to the right-hand side of the
equation (\ref{eq16}), and was therefore real. As expected, the
carrier frequency was observed with wide-band detectors. Yet, narrow
band detectors failed to see it again, just as in the experiments of
L\&R. Instead, both the Rb vapor and the Fabry-Perot cavity
``recognized'' the \emph{virtual} frequencies that would have been
required to produce the same net state with two beams. Technically
speaking, the right-hand state of the equation (\ref{eq16}) was
broken down into its left-hand side components, which is known as
Fourier decomposition. Gilra and collaborators also provided a very
sound interpretation of the interaction between modulated beams and
spectroscopic detectors, demonstrating that Fourier decomposition
can be described in a physically meaningful way without violating
the principles of classical mechanics. In particular, they explained
how amplitude modulation can prevent resonance at the carrier
frequency and enable it for down-converted components in narrow-band
filters. As a corollary, it does not matter if the complex modulated
state is produced by mixing two input frequencies, or by suitably
modulating a single input frequency. The products of the two
situations are experimentally (and not just mathematically)
indistinguishable.

As mentioned at the beginning of this section, the superposition of
an ideal red beam with an ideal yellow beam appears exclusively
compatible with the principle of non-interference. Yet, this
perception is based on partial knowledge. For a proper conclusion,
it helps to know that an ideal orange-colored laser beam can be
induced to split into a red beam and a yellow beam just by
modulating its amplitude, with suitable phase delays between the
beat frequencies of adjacent points in the cross-section
(Fig.~\ref{fig:freq}B). When the superposed state is generated by two
beams, it seems ``clear'' that they just pass through without
interaction. Yet, when the same state is produced by a single beam,
it is just as ``clear'' that unperturbed passage is impossible. The
two net states are mathematically identical by design, and every
detectable property of the output beams is going to be similar in
both contexts. Accordingly, the evidence is not exclusively
compatible with a single interpretation. A choice must be made on
the basis of independent considerations. The interference principle
entails a unified model, by treating the net states as if they were
physically identical in both cases. Incompatible oscillations cannot
coexist at single points in this model. In contrast, the concept of
non-interference entails that we are dealing with two different
mechanisms of propagation, just like in the case of polarized beams.
When two input beams are used, the net state is ``really'' made of
two independent beams. When a single input beam is modulated, the
net state is ``really'' monochromatic. Though, we cannot tell the
two situations apart by doing measurements on the superposed or the
output states. The only way to distinguish the two scenarios is by
learning the prior history of each context. Remarkably, this extra
information cannot influence any prediction about the future state
of the output beams. Moreover, the limited claim that beams can
\emph{sometimes} pass through each other unperturbed is not
necessarily true, even when the presence of two input beams is
verified. The evidence is equally compatible with the interference
principle, which excludes this possibility.

\subsection{Beams with different directions of propagation}

\begin{figure*}
\includegraphics{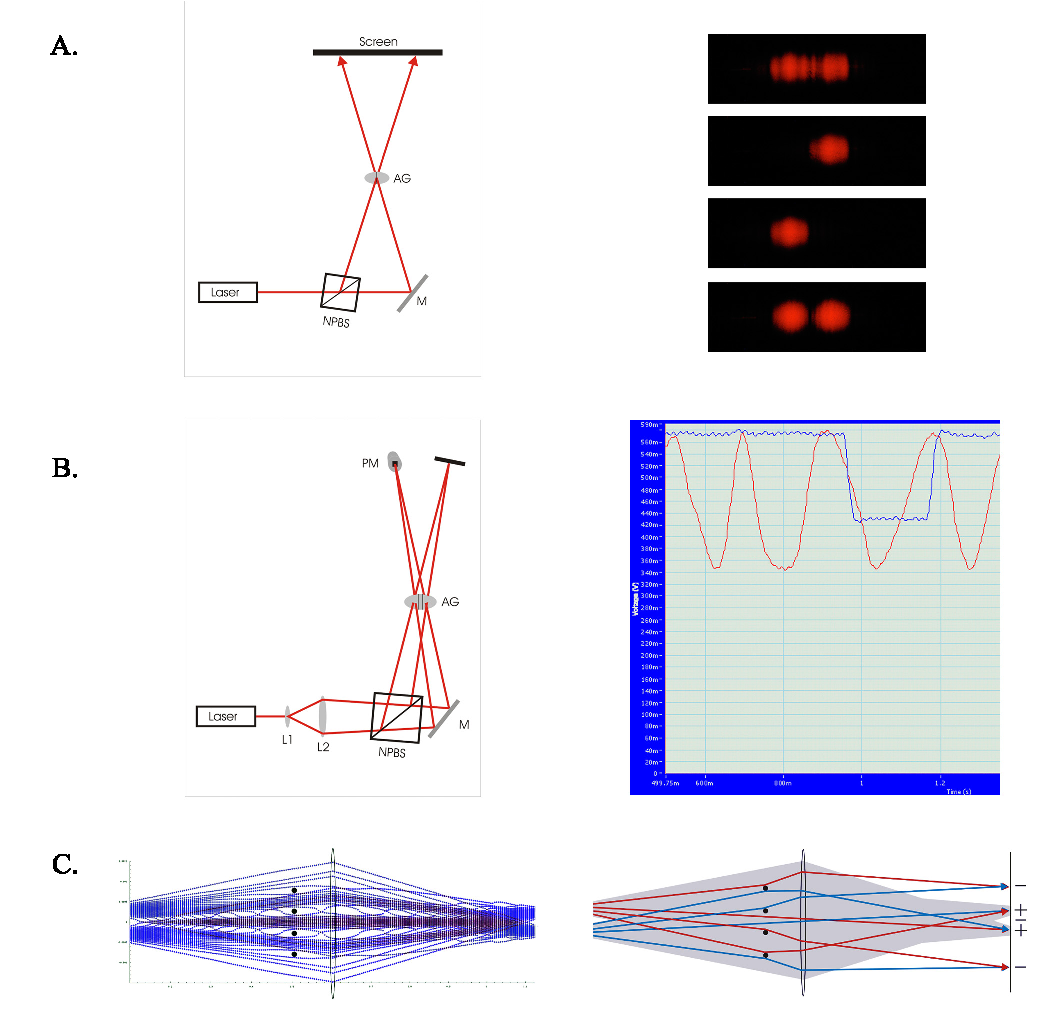}
\caption{\label{fig:afshar} (Color online) \textbf{Several illustrations
of the Afshar effect.} A) Two coherent laser beams intersect at a
very narrow angle (figure not to scale). The aperture AG contains a
single strand of human hair, stretched in the vertical direction. If
the obstacle is positioned in the center of a bright fringe,
excessive diffraction washes away the separation of the beams (top
image on the right) and the strand glows brightly (not shown). There
is less diffraction if the beams are open one by one (middle two
images). Diffraction becomes negligible if the hairline is placed in
the center of a dark fringe (bottom image). B) Similar set-up with
convergent beams. This allows for better beam separation and wider
fringes in the interference volume. Two 40ga wires are placed across
the aperture AG (each in a dark fringe). The power meter PM records
the irradiance of a single beam, after separation. The blue trace
(right image) shows a square drop in irradiance, observed when the
unmeasured beam was momentarily blocked, preventing interference.
The red trace shows a continuous variation of irradiance, recorded
when fringes drifted across the obstacles due to the displacement of
the mirror M at sub-wavelength intervals. Peak irradiance was
recorded when the wires were at the center of dark fringes. Notice
that the obstacles induced a larger drop in irradiance at the center
of bright fringes (red trace), compared to the case of
non-interference (blue trace). C) Estimated net energy distribution
in the x-z plane of the original Afshar set-up. Each blue dot
represents the center of a volume that contains 4\% of the input
energy of one beam. In ideal conditions, the separated beams look
the same with or without the wires in dark fringes (black circles).
Yet, the presence of such obstacles makes it impossible to describe
output beams as identical to input beams. Even if the underlying
propagation of radiation is assumed to be rectilinear, without
flowing around the obstacles, output beams have to be described in
terms of interference effects, as suggested by the drawing on the
right.}
\end{figure*}

Finally, let us consider a simple experiment, in which two coherent
beams are collimated and allowed to intersect at a very narrow
angle, producing the appearance of fringes in the interference
volume. In the absence of any disturbance, the beams will eventually
separate from each other and become clearly distinguishable.
According to the principle of non-interference, the fringes in the
interference volume are illusory. Light from each beam is uniformly
distributed in the cross-section, but the amplitudes of each
component add up constructively or destructively. The rays from each
source of light simply pass through each other. Next, suppose that
the interference volume is blocked with a screen, and only a fringe
is allowed to pass through a slit. The size of the slit is carefully
selected and the screen is positioned such that the edges of the
slit are in the centers of consecutive dark fringes. As far as
appearances are concerned, the edges of the slit do not produce any
scattering - they are in dark regions. Yet, the light that passes
through does not separate clearly into two beams - we are at the
boundary of the Rayleigh criterion. If we now repeat the procedure
for each bright fringe, it will become obvious that none of the
fringes is able to separate into two beams. However, if we put very
narrow obstacles at the center of each dark fringe, as if the edges
of a slit are still present there, and open all the fringes at the
same time (essentially, each bright fringe goes through its own wide
slit), we see very good beam separation, as if there are no
obstacles at all in the path of the beam. This is the well-known
Afshar modification of the double-slit experiment \cite{afs07,
afs05}. It has been hotly debated and there are numerous equivalent
ways to interpret it \cite{steu07, kas05, afs06, flo07, kol07,
geo07, flo08, jac08, ang09, flo10, dre11}. Some of our own results
on this topic are summarized in Fig.~\ref{fig:afshar}.

For the purpose of this discussion, we must distinguish only two
main interpretive approaches. On the one hand, we could assume that
the interference principle is real. In this case, we must also
assume that there is no energy in the dark fringes, because all of
it is channeled through the bright fringes. This explains why there
is no diffraction when both beams are open, even though significant
diffraction is visible when they are open one at a time. The beams
are expected to separate well with or without the obstacles, because
the underlying physical mechanism is the same in both situations.
Furthermore, beam separation is only possible when all the bright
fringes are open, because they have to overlap with each other for
this effect. The net state contains all the energy in bright
fringes, and it evolves into a pattern with two beam-like
projections, as predicted by the Huygens-Fresnel formalism. On the
other hand, we could also assume that non-interference is real. In
this case, the explanation is a little more intricate, but still
plausible \cite{steu07}. If the reality of interference is denied,
we must assume that diffraction at the obstacles for both beams
together is the same as in the case when the beams are open one by
one. This means that the input beams no longer pass through the
volume of superposition unperturbed. If their diffracted projections
on a remote screen are summed up, fringe by fringe, taking into
account the phase and amplitude of each component at every point of
detection, it will turn out that the net state adds up to zero
almost everywhere, effectively erasing the appearance of scattering.
At the same time, there will be two regions where the components
will add up constructively, producing the appearance of two
separated beams. If we measure the power of these two ``bright
fringes'', it will be nearly equal to the power of the input beams,
as if the energy has flown around the obstacles in the volume of
overlap. In other words, the two output beams must be interpreted as
illusory, because the actual radiation is widely scattered in space.
The identity relationship between input and output beams is denied,
just like in the case of real interference, but for a different
reason. When there are no obstacles in the dark fringes, beams are
really assumed to pass through. When the obstacles are present, they
are really assumed to scatter widely, even though the apparent
projections are identical in both cases. It is somewhat ironic that
the motivation behind non-interference models was to avoid the
complex calculations that are associated with Huygens' Principle,
yet the same model had to be invoked in order to explain the outcome
of the Afshar experiment. Not only is non-interference conceptually
challenging, it turns out that even its methodological advantages
have a limited scope.

To make things even more complicated for this model, the nature of
output projections is undefined even if there are no obstacles in
the dark fringes. As mentioned already in the preceding sections of
this chapter, the net state of two superposed beams can be mimicked
by modulating a single input beam. In the case of intersecting
coherent beams, this has been demonstrated with holographic tools
\cite{tho78, roy76}. When a single reference beam illuminates the
hologram of two superposed beams, it can faithfully replicate their
properties. Behind the holographic plate, it is possible to observe
interference fringes and even beam separation, as if two input beams
are passing through. In actuality, the holographic projection is
made by light scattered from a single input beam. The plate works as
a filter that controls the amplitude and phase profile at every
point in the cross-section of the projection. As predicted by the
Huygens-Fresnel model of wave propagation, the net state at any
front line is sufficient to reproduce the subsequent dynamics,
regardless of the conditions that produced it. In some contexts, it
is generated by two superposed beams. In others, it can be
engendered by modulating a single beam. This phenomenon shows that
net states are not just hypothetically real. In many cases, they are
undeniably so. Yet, the principle of non-interference forces us to
invent two different theories, even though any
observable property of these states is identical in both scenarios.

In conclusion, we see that the process of interference can be fully
interpreted by two incompatible models. Both of them make the same
predictions and often rely on the same (or demonstrably equivalent)
mathematical expressions. The two models are only incompatible in
the interpretive dimension. The principle of non-interference has
become an integral part of classical physics, and it seems to be
widely taken for granted as a Newtonian process. As shown at the
beginning of this text, it is actually a non-classical mechanism,
requiring the assumption of non-locality in the analysis of both
motion and energy, and also leads to several conceptual
inconsistencies. We asked if these shortcomings were compensated by
some unexpected interpretive advantage when particular applications
were considered. After reviewing the properties of superposed beams
distinguishable by polarization, frequency and/or direction of
propagation, this sort of benefit was not found. Quite the opposite:
physical observations (and the equations that fit them) became
indeterminate in this context. Whenever a beam is measured, it is
impossible to tell if it is an actual single-mode projection, or a
mixture of two components with illusory energy content, even though
it is a classical state. The only way to know the ``truth'' is by
acquiring the full prior history of the projections, all the way to
the source. Yet, we cannot extract anything consequential out of
such information, because this interpretation does not entail any
new prediction. The subsequent behavior of the investigated beam is
going to be the same in any experimental setting, regardless of any
hypothetical distinction in its nature. In short, we need two
explanatory models for a single type of observations, even though it
is possible to interpret both scenarios with a single classical
mechanism. Moreover, each of the two models of non-interference is
more complicated than then single construct that replaces them if
interference is assumed to be real. As a corollary, non-interference
cannot survive Ockham's razor.

\section{Quantum implications}

The principle of linear superposition tells us that the net effect
of two waves on a single point is reducible to the individual
contribution of incident components. Though, it cannot help us
decide which states are real during overlap: the individual inputs
or the net output? We have a mathematical equality between the two
alternatives, as well as an experimental equivalence, as shown
above. Accordingly, it does not seem to matter which model is
chosen: when every detail is taken into account, the final
predictions are the same in each case. With this in mind, it is
worth asking: what difference does it make if we prefer one story or
the other? If we assume that waves do not interact, we get a
non-classical picture, as well as many complications, but it still
works. This has been the favored model for many generations of
scientists. Why do we have to reconsider it? The answer is that
interpretations are based on hypotheses about underlying
(microscopic) processes. In effect, they are theories about quantum
phenomena. The concept of wave non-interference was part of the
mainstream since at least the 19th century, before the development
of tools for testing such assumptions. Yet, quantum mechanics is now
widely believed to falsify the validity of classical mechanics at
the microscopic level. This means that the descriptions of classical
processes, such as interference, are not compatible with the known
microscopic phenomena. Ergo, prevailing \emph{interpretations} of
classical phenomena are incompatible with quantum observations. What
if we tried to interpret classical wave superposition with a truly
classical model, in which interference is real? Would we still get a
conflict with quantum mechanics?

Many quantum phenomena result from wave-like interactions, and they
are predicted on the basis of wave-function analysis. Remarkably,
the principle of superposition applies in this context as well,
except we have a non-local version. In classical physics,
superposition is described as a coincidence between two entities
with well-defined states. In quantum physics, we have single
entities that are described as undefined, because they occupy
multiple states at the same time \cite{dirac}. This can be
illustrated with coherent beams of light that have orthogonal
polarization. They can be obtained by splitting a single laser
projection in two, and then placing polarizing filters in each path
(\emph{e.g.}, one horizontal, and one vertical). At the classical
level, the beams are continuous. If they are forced to overlap in
phase, the net state is a beam with diagonal polarization (as shown
in Fig.~\ref{fig:mzi04}). So, the beam will be transmitted in full by a
polarizing beam-splitter with diagonal fast axis, if both paths are
open. In contrast, each component must send 50\% of its energy in
the reflected path, if it arrives at the analyzer alone. The same
type of behavior is observable at the quantum level. If the source
is attenuated, it is possible to observe discrete detection events.
Various control experiments can be used to show that detection
events are most likely produced by indivisible wave-packets
\cite{zei99}. Yet, these single wave-packets display all the
features of superposition. When both paths are open, they are always
detected in the transmitted path. When one path is open at a time,
half of them trigger detections in the reflected path. This shows
that single quanta display the properties of a net state that
requires two real components (or more) in the context of classical
mechanics.

The most remarkable feature of quantum distributions is their
similarity with classical detection patterns \cite{bohr76}. Discrete
photons are supposed to have a non-classical nature \emph{because}
they are able to reproduce the behavior of continuous beams. To be
exact, the problem is not that they are discrete. Rather, it is
their undefined nature. Unlike classical particles, they are
required to be in several states at the same time. Why are they
described like that? The answer is found in the details of classical
interpretations of superposition. When the governing assumption
dictates that waves do not interfere, the net state can only be
described as an illusion. This means that the diagonal polarization
in the preceding example is assumed to be produced by the combined
effect of two real states of polarization (horizontal and vertical).
In other words, the classical beam is in two states at the same
time. For this reason, a single photon in the state of the same beam
must also be described as if it was polarized in two planes at the
same time. Consequently, we have no choice but to describe a quantum
with non-local concepts: it has to move in two directions at the
same time, be in two places at the same time, and so on.
Nevertheless, the mainstream interpretation is not the only one that
works at the macroscopic level. As shown above, it also violates the
spirit of classical mechanics. In contrast, if we assume that
interference is real, then the net state must also be treated as
real. This means that the polarization of the output beam is really
diagonal in the preceding example, whether we are looking at
continuous or discrete states of light. As a corollary, the same
single photons can now be described as well-defined, without any
change in their quantitative description. Diagonal states of
polarization are mathematically equivalent to the sum of two
orthogonal states of polarization (assuming equal intensity and
phase coherence). This means that we get the same predictions,
regardless of the associated interpretive assumptions. On closer
inspection, it turns out that this conclusion can be generalized to
every quantum phenomenon where multiple states are found in
superposition. Wherever it is possible to describe a quantum as if
it was occupying multiple states at the same time, it is also
possible to say that it belongs to a system that occupies the single
net state. Ergo, quanta in ``Schr\"{o}dinger's cat'' states can
always be interpreted as entities with well-defined properties. This
does not mean that real cats can be in the net state of ``dead +
alive''. Detection events are never superposed (\emph{viz.}, the
measurement problem). It is the net state of a quantum that
determines the probability of generating one type of event or
another. In short, the nature of our interpretive conclusions at the
microscopic level is not determined by the details of our
observations, but rather by the interpretive choices at the
macroscopic level of analysis. If we switch to a classical
interpretation of superposition at the classical level, we also get
a classical interpretation at the quantum level.

A similar sensitivity to macroscopic preferences is demonstrable for
the principle of quantum complementarity. The latter was developed
by invoking explicitly the double-slit experiment \cite{bohr87,
schilp}, which makes it particularly relevant for this discussion.
When two beams of light intersect, they seem to go through each
other as if they never meet. The principle of non-interaction holds
that this is, indeed, the case. Interference fringes are detectable
in the volume of overlap, but they are described as mere
appearances. By implication, if microscopic bits of light were
detected one by one in the interference volume, they would be
resolved in their ``real'' state during overlap and shown to display
continuous distributions. These elements of light must continue
propagating undisturbed, with rectilinear trajectories, as suggested
by the method of ray tracing in geometrical optics, and carry path
knowledge past the interference volume. Unfortunately, quantum
experiments falsified this expectation. Populations of single quanta
were found to display interference fringes just like classical
beams, and also to separate into distinguishable beams after
crossing the volume of overlap. Obviously, this was a big problem,
and the solution was to introduce the principle of complementarity.
According to this new story, single quanta were presumed to produce
fringes only when directly observed. Otherwise, they were still
assumed to remain unperturbed and to carry path knowledge all the
way from the source to the detector. In other words, two
complementary realities were assumed to take place, associated with
two incompatible observations. The human choice between alternative
measurements determined which reality played out. Such an
interpretation seems necessary because it is impossible to describe
a plausible physical model in which path knowledge and interference
knowledge are simultaneously present. Nevertheless, it is possible
to describe a physical reality in which quantum interference and
beam separation are possible in the same course of events. This is
the scenario that follows naturally from the assumption that wave
interference is real. In this case, microscopic bits of light are
naturally expected to flow through the bright fringes (avoiding the
dark ones) and to separate into distinguishable projections
afterwards. Of course, path knowledge is no longer assured in this
case, and the output beams are only quantitatively similar to the
input beams. Still, this interpretation is able to explain quantum
phenomena without inconsistencies. Accordingly, any belief about
non-classical quantum behavior in this context depends on a
commitment to non-classical interpretations of classical wave
interference. The Copenhagen interpretation is only meaningful if
the principle of non-interference is taken for granted.

\begin{figure*}
\includegraphics{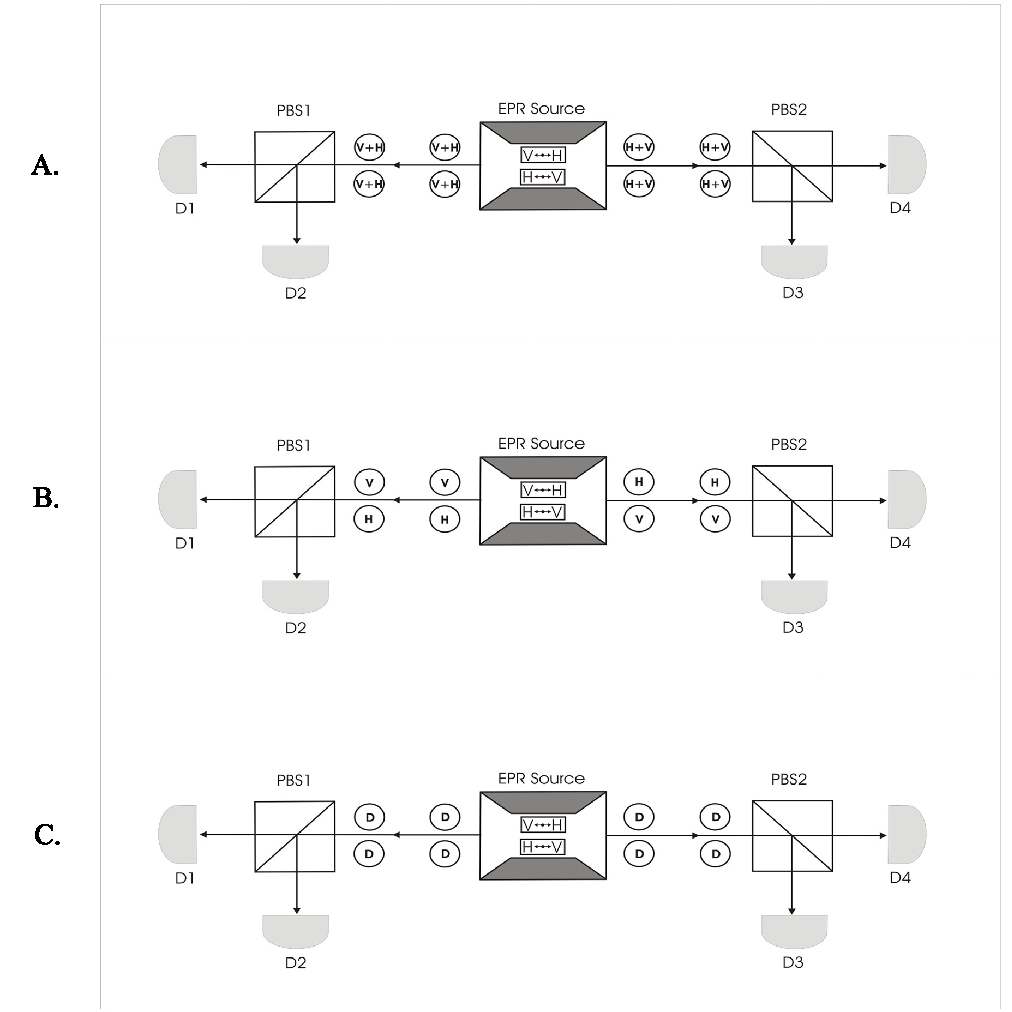}
\caption{\label{fig:epr} \textbf{Three interpretations of an EPR state.}
A non-linear source produces low intensity beams of entangled
photons pairs. Every photon in the right arm of the set-up has a
``twin'' with orthogonal polarization in the left arm. For
simplicity, the photons in the same arm are assumed to be randomly
distributed between two states of polarization only (vertical and
horizontal). Also, we assume that even numbers of photons are
emitted in phase by the source at any point in time, in order to
bring out the differences between interpretations. Every photon is
either vertical or horizontal at the source, but the subsequent
state can only be determined after measurement. (A): The Copenhagen
Interpretation holds that unmeasured indistinguishable photons
cannot have well-defined states of polarization. Each propagating
photon is assumed to be simultaneously in both states. This property
determines the outcome of subsequent measurements for any angle of
alignment of PBS1 or PBS2. Bell's Inequality can be violated in
predictable situations. (B) and (C): Classical models hold that
single photons have well-defined states at all times, but they can
evolve in two different ways. In (B), there is no interference and
photons preserve their input states during superposition. Detectors
are assumed to resolve individual photons, rather than macroscopic
states as a whole. Bell's Inequality cannot be violated, because the
final distributions are produced by unperturbed input states. In
(C), interference is assumed to be real, and photons switch from
their input states to the plane that corresponds to the net state.
In this case, every single photon is assumed to be polarized in the
diagonal plane. The distribution of detection events is no longer
produced by input states, but rather by the value of the net state,
as in (A). Bell's inequality can be violated. In this context, EPR
correlations serve as indicators of real interference. It is not
necessary to invoke non-locality for their explanation.}
\end{figure*}

These examples suggest that quantum mechanics can be interpreted as
a theory with well-defined local objects. Such an outcome is
generally perceived as impossible, because of the EPR paradox
\cite{bell}. Entangled quanta have identical properties. Therefore,
one can obtain exact information about incompatible variables by
measuring several populations of entangled quanta with different
devices. This appears to imply that all of those properties are
well-defined prior to measurement, in violation of the uncertainty
principle. Yet, well defined entities in such states must
necessarily obey Bell's inequality. The unexpected finding,
supported by numerous careful experiments (starting with
\cite{aspe1, aspe2, aspe3}), is that quanta violate this rule. The
paradox is that we need entangled systems to reveal the apparent
reality of complementary properties, but the same entities violate
Bell's inequality and thereby refute the implication of reality.
That is why it does not work to assume that well-defined population
components exist in the absence of measurement. For example, when
there is a beam with two components of polarization -- vertical and
horizontal -- it is tempting to assume that some photons are
vertical while others are horizontal, but this would entail a
contradiction with empirical observations. The correct way to
describe single quanta of light is by assuming that each of them is
in both states of polarization at the same time. This is what makes
quantum mechanics so hard to comprehend: a single entity must
oscillate in two incompatible directions at the same time, like the
classical beam as a whole. Then again,  what happens to a classical
object, such as a rope, when it is agitated in two orthogonal planes
at the same time? It must respond to both actions by oscillating in
the diagonal plane (assuming phase coherence). From a mathematical
point of view, the rope is in a state of oscillation that is equal
to the two input states. Hence, it does contain both states at the
same time. However, the physical reality is that the rope cannot
oscillate in opposite directions. It must remain well-defined by
oscillating in the net state. If two identical ropes were measured
simultaneously -- one in the vertical plane, and the other in the
horizontal plane -- it might seem as though the rope had real states
of vibration in each plane. Yet, as a physical description, such a
conclusion would be wrong. A pair of ropes would also violate Bell's
inequality in terms of oscillation components, because it is the net
state that determines the measurement outcomes. Accordingly, a
classical photon can also be described as being in a superposition
of two states of polarization, if in fact it is polarized in the net
state (Fig.~\ref{fig:epr}). The EPR paradox was produced by the
assumption that component states of superposition are real, while
the net state is not. Such a belief can only seem necessary if the
principle of non-interaction of waves is taken for granted. Even so,
this is purely an interpretive preference. If
$\bm{A}+\bm{B}=\bm{C}$, then the two equal expressions are
interchangeable. It is just as valid, from a quantitative point of
view, to assume that the component states are real, or that the net
state is real. Accordingly, it is possible to switch to the
principle of real interference and avoid the paradox without losing
predictive power.

We wish to emphasize that classical interference and quantum
interference are formally compatible with each other, as implied by
Bohr's correspondence principle \cite{bohr76}. The gap between
macroscopic and microscopic phenomena is not quantitative. We can
use the same equations to predict the details of interference
fringes in both cases, for large $N$. Continuous beams and discrete
populations of photons generate the same types of distributions on
detector screens. The gap is rather qualitative: we cannot use the
mainstream classical interpretations at the quantum level without
running into complications. We are forced to invent incompatible
realities and non-local entities in order to fill the conceptual
void that is created by the adopted classical \emph{interpretive}
models. Consequently, the experimental record of quantum mechanics
does not entail the collapse of classical formal analysis. It only
entails a violation of the assumptions that are associated with
leading interpretations of classical wave superposition. More
importantly, as shown in the preceding chapters, these assumptions
are already in conflict with the main principles of Newtonian
physics. An equivalent interpretation that does not create such
difficulties is readily available. We only have to assume that
interference is real. When two waves overlap, the net state is real,
while the input components lose their physical identity. Of course,
this argument does not prove that the Copenhagen interpretation is
wrong. It only shows that it cannot be true with necessity.
Classical alternatives work at least as well.

\section{Concluding remarks}

Linear superposition and the non-interaction principle have a long
common history. In many contexts, it is very convenient to assume
that waves can pass through each other unperturbed, with the
benefits of simplified algebraic and geometrical representations.
Unfortunately, the idea of linear superposition without energy
redistribution entails a foundational inconsistency in classical
mechanics. If we take it for granted, we need to make several
non-Newtonian assumptions about underlying processes (\emph{e.g.},
particles can move in two directions at the same time and transmit
momentum without moving at all). Also, we need to use the wrong
values for the wave amplitudes in order to make correct predictions
(or make additional non-Newtonian assumptions, in order to argue
that those amplitude values are correct). In addition, we get the
contradictory observation of energy redistribution (despite the
starting motivation to avoid it), having to explain it with
non-local mechanisms. Furthermore, we end up with indeterminate
conclusions in many contexts of observation, because different event
histories lead to similar observations. Identical phenomena,
described by identical equations, have to be interpreted in
different ways, without any compensatory practical benefit. Finally,
the microscopic assumptions of this model are contradicted by the
quantum-level observations, imposing the necessity to formulate new
interpretive models (\emph{e.g.}, the Copenhagen interpretation) for
this level of analysis. In order to hold on to the assumption of
non-interference, we have to assume that Nature is governed by an
inconsistent mixture of laws. This list of difficulties is probably
surprising, considering our real life experiences. If waves are
examined in a pond, for example, it seems obvious that they pass
through each other. Notwithstanding, the Sun also seems to move
around the Earth every day. If we take that experience for granted,
it is very difficult to unify our Earthly observations with the
totality of our knowledge. Likewise, if we take it for granted that
waves are transparent to each other, there is little hope of
unifying classical mechanics with quantum mechanics.

The main principles of non-classical physics belong to a network of
mutually reinforcing assumptions. The latter cannot be proven to be
correct or wrong, but their intuitive appeal is heavily influenced
by the perceived validity of the non-interaction principle during
classical linear superposition. As shown above, quantum
superposition and quantum complementarity do not seem plausible
otherwise. The strength of these assumptions is also backed by their
compatibility with existing formal models that work with
unprecedented accuracy. If the equations work so well, how could the
interpretations be wrong? Would it not be necessary to invent a
better formalism for a better interpretation? The answer is that
classical linear superposition provides a common formal backbone for
two incompatible interpretations. As soon as we change our story
about wave superposition, the same equations in quantum mechanics
acquire a radically different meaning. The crucial distinction
between ``classical'' and ``non-classical'' physics is not found at
the boundary between macroscopic and microscopic phenomena. Instead,
it is discovered at every level of analysis, where the terms of the
most simple equations are mistaken for the most relevant physical
elements. More importantly, we do not have to prove that
non-classical models are impossible. It is sufficient to acknowledge
that classical models are equally compatible with the same equations
and experimental data. In our opinion, it should be possible to
reclaim quantum mechanics -- in its present form -- as a classical
theory. Our arguments have only covered the direct implications of
linear wave superposition, but they justify a wider inquiry into the
ontology of quantum mechanics.

\begin{acknowledgments}
At various stages of its evolution, the argument of this paper was
influenced by feed-back from Yanhua Shih, Yoon-ho Kim, Giuliano
Scarcelli, Robert Boyd, Paul Kwiat, Y.S. Kim, Stephen Walborn, Ana
Maria Cetto, Alain Aspect, C. S. Unnikrishnan, Tim Maudlin, Sheldon
Goldstein, Arkady Plotnitsky, Serafino Cerulli-Irelli, Marlan
Scully, Marcus Appleby, Ernst Knoesel, Eduardo Flores, Shahriar
Afshar, Ken Wharton, Giorgio Kaniadakis, Frank Zimmermann, Piers
Coleman, James Franson, John Howell, Gregg Jaeger, Ryan Bennink,
Andrei Khrennikov, Guillaume Adenier, Al Kracklauer, Andrew
Meulenberg, Robert Hudgins, and Chandrasekhar Roychoudhuri. The
project was made possible by private donations to \emph{Open World
Research}.
\end{acknowledgments}


%

\end{document}